\documentclass[fleqn,usenatbib]{mnras}

\usepackage{newtxtext,newtxmath}

\usepackage[T1]{fontenc}
\usepackage{ae,aecompl}

\usepackage{graphicx}	
\usepackage{amsmath}	
\usepackage{amssymb}	
\usepackage{siunitx}
\usepackage{xspace}
\usepackage[flushleft]{threeparttable} 
\usepackage{dblfloatfix} 

\usepackage{xcolor}
\usepackage{comment}

\title[New $\beta$~Cep pulsators discovered with \textit{K2}]{New $\beta$~Cep pulsators discovered with \textit{K2} space photometry}

\author[S. Burssens et al.]{S. Burssens$^{1}$\thanks{E-mail: \href{mailto:siemen.burssens@kuleuven.be}{siemen.burssens@kuleuven.be}},
D. M.~Bowman$^{1}$,
C.~Aerts$^{1,2}$,
M. G.~Pedersen$^{1}$,
E. Moravveji$^{1}$,
\newauthor{B. Buysschaert$^{1,3}$}
\\
$^{1}$ Instituut voor Sterrenkunde, KU Leuven, Celestijnenlaan 200D, 3001 Leuven, Belgium\\
$^{2}$ Department of Astrophysics, IMAPP, Radboud University Nijmegen, 6500 GL Nijmegen, The Netherlands\\
$^{3}$ LESIA, Observatoire de Paris, PSL Research University, CNRS, Sorbonne Universit\'{e}s, UPMC Univ. Paris 06, Univ. Paris Diderot, \\ Sorbonne Paris Cit\'{e}, 5 place Jules Janssen, F-92195 Meudon, France \\
}

\date{Accepted XXX. Received YYY; in original form ZZZ}

\pubyear{2019}

\begin{document}
\label{firstpage}
\pagerange{\pageref{firstpage}--\pageref{lastpage}}
\maketitle

\begin{abstract}
We present the discovery of three new $\beta$~Cep pulsators, three new pulsators with frequency groupings, and frequency patterns in a B3Ib star, all of which show pulsations with frequencies as high as about 17 d$^{-1}$, with \textit{K2} space mission photometry. Based on a Fourier analysis and iterative pre-whitening we present a classification and evaluate the potential for asteroseismic modelling. We include the lists of pulsation frequencies for three new $\beta$~Cep pulsators, CD-28~12286, CD-27~10876, LS~3978, and additional pulsation mode frequencies for the known $\beta$~Cep pulsator HD~164741. In addition we characterise the regular frequency spacing found in the new pulsator HD~169173, and discuss its origin. We place the newly discovered variables in a colour-magnitude diagram using parallaxes from Gaia-DR2, showcasing their approximate location in the massive star domain. The identified frequency lists of these multiperiodic pulsators are a good starting point for future forward seismic modelling, after identification of at least one pulsation frequency from high-resolution time series spectroscopy and/or multi-colour photometry. 
\end{abstract}




\section{Introduction}\label{section: introduction}

Despite massive stars not being as common as lower mass stars, they are the metal factories of the Universe enriching their surroundings by means of stellar winds and supernovae explosions. Massive star physics such as fast rotation, mass-loss and binarity complicate stellar modelling and evolutionary predictions \citep{Maeder2009}. Consequently the interior physics of massive stars remains poorly constrained: specifically the shape of core overshooting \citep{Moravveji2015, Moravveji2016b, Salaris2017,Pedersen2018}, angular momentum transport \citep{Aerts2019a}, and chemical mixing are all phenomena of which we have limited observationally motivated understanding. In the last decade increasing efforts have been made to understand these stars by means of asteroseismology. This provides a powerful way to study both the global and local physics of the star, see \citet{Aerts2010} for an extended overview of this method. 

The advent of space missions such as \textit{CoRoT} \citep{Auvergne2009}, \textit{Kepler} \citep{Borucki2010}, \textit{K2} \citep{Howell2014}, and more recently \textit{TESS} \citep{Ricker2014}, heralded a new age in the asteroseismology of OB-type massive stars. The high duty cycles and long time bases uninterrupted space photometry are the necessary data properties for the asteroseismology community. Pulsation frequencies can be extracted via iterative pre-whitening (see \citealt{Lenz2004, Degroote2009a, Papics2012a, Bowman2017}), and if regularities in the extracted pulsation mode frequencies are detected, inference is possible of the interior of the star \citep{VanReeth2016, Papics2017, VanReeth2018}. Pulsations in early-type main sequence stars generally occur in two frequency regimes: the low-frequency gravity modes and the high-frequency pressure modes. Gravity (g) modes are standing waves for which the dominant restoring force is buoyancy. These probe the near-core region of a star. Pressure (p) modes are standing sound waves for which the restoring force is the pressure force, and predominantly probe the stellar envelope. Each wave is characterised by three numbers: $n$ for the radial order, $\ell$ for the spherical degree, and $m$ for the azimuthal order \citep{Aerts2010}. In the case of high radial orders ($|n| >> \ell$) in a chemically homogeneous non-rotating star, p- and g-modes are equally spaced in frequency and period, respectively, according to the asymptotic theory of stellar oscillations \citep{Tassoul1980}. Deviations caused by chemical composition ($\mu$-) gradients or rotation, lead to dips (for the former) or a slope (for the latter) in the g-mode period spacing pattern \citep{Miglio2008, Bouabid2013, VanReeth2015b, Ouazzani2017}. 

A recent example of using forward seismic modelling that exploited g-mode period spacing patterns in a Slowly Pulsating B star (SPB) is \citet{Moravveji2016b}, who argued that the g-modes in KIC~7760680 indicate moderate rotation, convective core overshooting as well as diffusive mixing in the radiatively stratified layers of the envelopes. Similarly, by modelling observed g-mode spacing, \citet{Papics2017} and \citet{Szewczuk2018} were able to measure the near-core rotation in a selection of SPBs, and constrain interior chemical mixing using {\it Kepler} mission data. As discussed in detail by \citet{Pedersen2018}, modelling g-mode period spacing patterns can yield information about the specific shape and size of the overshooting. In \citet{Buysschaert2018b} it was shown that convective overshoot suppression by large-scale magnetic fields can be inferred by modelling g-modes, for HD~43317.
The power of using g-mode period spacing patterns for stellar modelling for massive stars, is extensively discussed in \citet{Aerts2018b}.

Additionally, \textit{CoRoT} data have contributed to the advancement of asteroseismology of upper main sequence pulsators. The origin of the observed oscillations in the only known $\beta$~Cephei ($\beta$~Cep) star observed by \textit{CoRoT}, V1449~Aquilae, has been extensively discussed in \citet{Belkacem2009}, \citet{Degroote2009a} and \citet{Aerts2011}. A few massive O-stars were also observed, \citet{Degroote2010b} noted regular frequency spacings in the young O-type binary HD~46149 and attributed it to stochastic excitation in the subsurface convective zones. A detailed asteroseismic study by \citet{Briquet2011} on the O9V star HD~46202 demonstrated that the excitation of pulsations in massive stars is still not globally understood. A similar conclusion was obtained by \citet{Blomme2011} for three hot O-type stars where the frequency spectrum is dominated by astrophysical red noise, which was later suggested to be caused by internal gravity waves by \citet{Aerts2015b} and \citet{Bowman2019a}.
The complex pulsations in rapidly rotating Be stars also became accessible with \textit{CoRoT} \citep{Neiner2009, Diago2009, Gutierrez2009,  Neiner2012}, similarly demonstrating that more theoretical work is needed, in understanding the diverse variability observed in early-type stars.

Here we focus on the detection of new heat-driven OB-type variable stars, which are stars whose oscillations are driven by the $\kappa$-mechanism \citep{Kiriakidis1992, Moskalik1992, Gautschy1993, Dziembowski1993a, Dziembowski1993b}, discovered in the \textit{K2} space mission data. There are two main types of pulsator amongst early-type stars: slowly pulsating B (SPB) stars SPBs \citep{Waelkens1991}, which have an approximate range of log~$T_{\rm eff}\in$ ~[4.05, 4.35] and log~$L/\text{L}_{\sun}\in$~[2.0, 4.0], and the hotter, more massive $\beta$~Cep pulsators (\citealt{Frost1902}, and see the overview given by \citealt{Stankov2005}) which have an approximate range of log~$T_{\rm eff}\in$~[4.25, 4.50] and log~$L/\text{L}_{\sun}\in$~[3.2, 5.0].

Although the oscillations are driven by the same mechanism, the character of the pulsations is different: high-order g-modes are typically found in SPB stars, and low-order p- and g-modes in $\beta$~Cep stars. The particular opacity bumps responsible for mode excitation in OB-type stars have been found to be the partial ionisation layers of iron-group metals. \citep{Dziembowski1993a, Dziembowski1993b, Pamyatnykh1999}. There are many factors influencing the driving zones, such as the chemical composition, atomic diffusion, macroscopic mixing and stellar metallicity. Attempts at defining instability regions are in general agreement --- \citet{Pamyatnykh1999} presented a discussion on the instability region of the $\kappa$-mechanism in the upper main sequence. \citet{Deng2001} argued that core overshoot has little effect on the iron bump opacity shape, based on linear stability analyses of stellar models. \citet{Moravveji2016a} investigated the effects of enhancing the iron and nickel opacities in early-type stars and found the instability regions were wider. More recently, \citet{Szewczuk2017} studied the effect of rotation on the instability region. One of their results is that rotation has a major effect on the extent of the instability domains of SPBs in comparison with core overshooting, opacity data, initial metallicity and initial hydrogen abundance.

The sample of main sequence massive stars with adequate photometric data for asteroseismic modelling studies remains limited. The \textit{Kepler} field of view (FOV) was purposefully chosen to avoid star-forming regions in which OB-type stars are typically found. Therefore the number of these stars to study was limited \citep{Balona2011, McNamara2012}. \textit{CoRoT} while successful, only provided a handful of stars for which detailed asteroseismic analyses were possible. 
On the other hand, \textit{K2} \citep{Bowman2019b}, and more recently \textit{TESS} (see \citealt{Pedersen2019a} for diverse variability of OB stars in \textit{TESS} sectors 1 and 2), have begun to remedy this lack of continuous, long-term, high-precision photometric data for massive stars.

\section{Method}\label{section: method}

\subsection{The \textit{K2} space mission}
In this work we focus on the data available from the \textit{K2} space mission \citep{Howell2014}, the re-purposed \textit{Kepler} mission. The radiation pressure of the Sun was used to minimise the use of on-board thruster fuel needed for stability. As such, an individual portion of the ecliptic was monitored for up to 90~d, until it was necessary to rotate the spacecraft to prevent the Sun from entering the telescope's field of view. Each of these monitoring campaigns viewed a new field, allowing for the measurement of objects that were not visible in the original \textit{Kepler} field.  Of particular interest for massive star research were Campaigns 2, 9 and 11 as these included star-forming regions around the galactic bulge, and therefore contained a variety of massive stars. The suitability of \textit{K2} for asteroseismology of OB-type stars has been demonstrated in the past, including studies for O stars in Campaign 0 \citep{Buysschaert2015}, chemically peculiar and magnetic stars \citep{Buysschaert2018a, Bowman2018}, evolved early-type stars \citep{Aerts2018a}, and B-type stars \citep{Balona2016, White2017, Aerts2018a}.

\begin{table*}
\caption{Sample of OB stars in \textit{K2} fields showing independent coherent pulsations in the high-frequency regime. References are made to the spectral type classification or to previous photometric studies of the object.}\label{table:Star_List}   
\begin{threeparttable}
\centering          
\begin{tabular}{c c c c c c c c c c}     
\hline
 Campaign \# & EPIC~ID& Star name & RA & Dec & V$_{\text{mag}}$ & Spectral type & $\Delta$ T (d)& N$_{\text{obs}}$& References\\ 
 \hline

&&&&&&&&& \\
0 & 202060092&	HD~256035&		06 22 58.24&	+22 51 46.17&	9.21& O9V:p & 36.19 & 1679& 1,2,3\\
&&&&&&&&& \\
2 &202691120&	CD-28 12286&		16 41 01.98&	-28 43 20.86&	10.32&	OB$^{-}$& 76.94 & 1904 & 4\\
&202929357&	CD-27 10876&		16 16 15.05&	-27 51 19.91&	10.95&		OB$^{-}$    & 64.25& 2302 & 4\\
&&&&&&&&& \\
9 &223832867&	HD~164741	&	18 03 44.08&	-25 18 45.08&	9.19	&	B2Ib/II& 71.34& 3055 & 5,6,7 \\
& 227552090	&HD~169173 	& 	18 24 08.16 &	-17 51 48.73 &	9.93	&	B3Ib & 71.34 & 2964& 5,6\\
&&&&&&&&& \\
11 & 233986359&	HD~156491 &		17 18 38.77 &	-21 35 21.81& 	9.54	&	B3III & 71.59 & 2957 & 5 \\
&235094159	&LS 3978 		&17 13 01.32 	&-24 12 18.88 	&12.68	&	B2III & 72.51& 3012 & 8  \\
& 235151005&	HD~157548 &		17 24 51.63 &	-23 55 33.94 &	8.8	&	B5III & 72.08 & 2971& 5 \\

\hline
\end{tabular}
\begin{tablenotes}
\small
\item \textbf{References}: 1) \citet{Negueruela2004}, 2) \citet{Buysschaert2015}, 3) \citet{Balona2016},  4) \citet{Drilling1995}, 5) \citet{Houk1988}, 6) \citet{Pigulski2008}, 7) \citet{Hohle2010}, 8) \citet{Vijapurkar1993}.
\end{tablenotes}

\end{threeparttable}
\end{table*}

\subsection{Light curve extraction}\label{subsection: LE} 
\begin{figure}
	\includegraphics[width=\columnwidth]{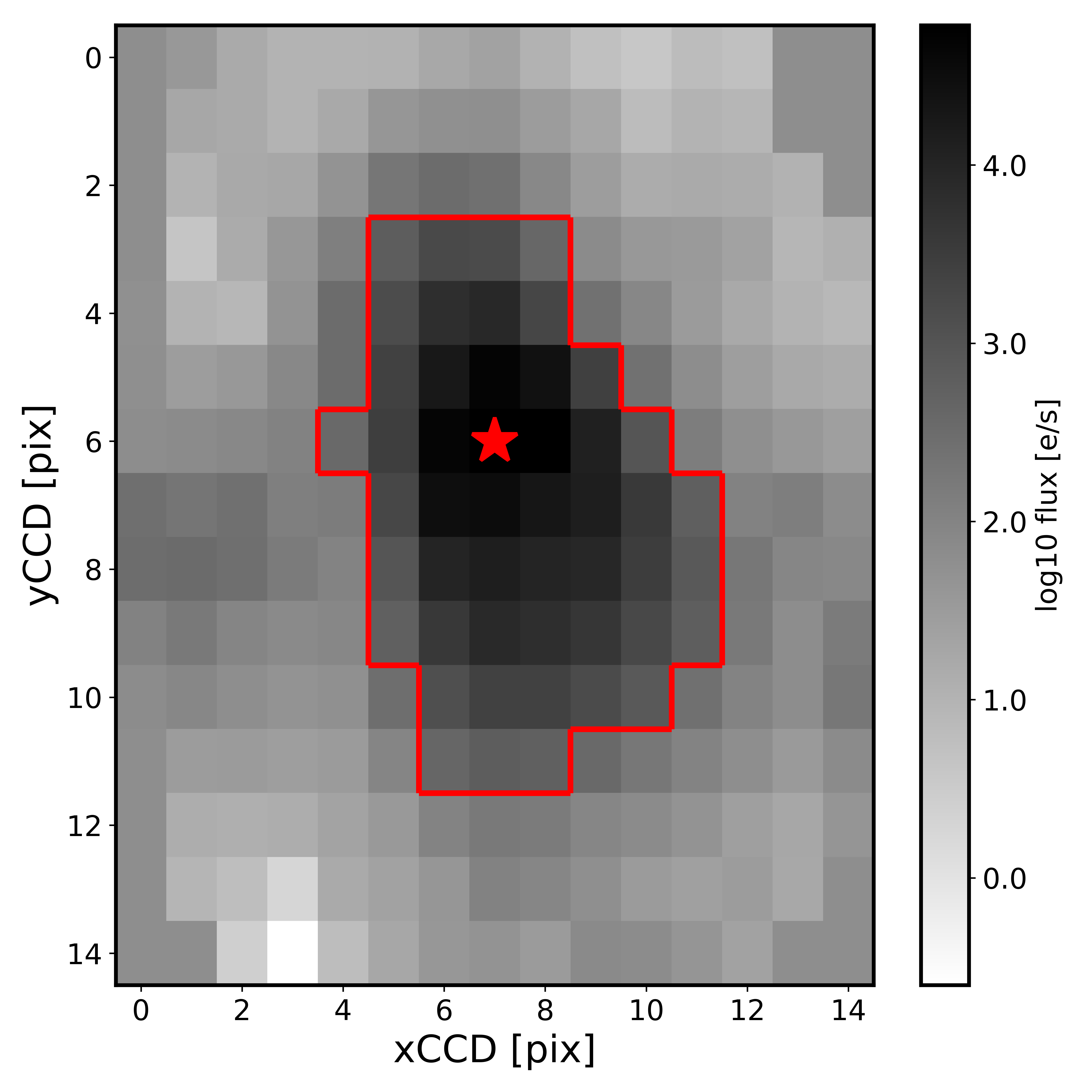}
    \caption{Example of a pixel mask for EPIC~202929357. The abscissa and ordinate indicate the CCD X and Y pixel location. The grey-scale bar shows the total flux. The red contour indicates the definition of the mask. The central star symbol is the centroid of the star and is where the semi-random walk is initiated.}
    \label{fig:example_pixel_mask}
\end{figure}

Target pixel data for targets observed by \textit{K2} can be retrieved from  MAST\footnote{Mikulski Archive for Space Telescopes (MAST), \href{https://archive.stsci.edu/}{https://archive.stsci.edu/}}. 
\textit{K2} data contains systemics due to the pointing jitter of the spacecraft \citep{Vanderburg2014, Aigrain2016}. Thruster firings occurred every \si{\sim6}~h to correct the satellite pointing. To extract the photometry from the target pixel frames we follow the  methodology of \citet{Buysschaert2015, Buysschaert2018a}. An optimal pixel mask is defined semi-automatically, including all pixels containing flux from the target by means of a semi-random walk whilst carefully avoiding possible sources of contamination from nearby or background stars. An example of such a mask is given in Fig.~\ref{fig:example_pixel_mask}, for the newly discovered $\beta$~Cep star EPIC~202929357. The flux counts within the optimum mask are summed at a given time stamp to produce a light curve.

We subsequently applied the \textit{k2sc} software package \citep{Aigrain2016}. This pipeline uses Gaussian Process regression to model both position (pointing systematics) and temporal (intrinsic to the star and any other non-positional systematics) variability simultaneously. White noise is handled as a third component. Gaussian Process regression is well-suited for this purpose as it does not require any prior information of the stellar variability, see \citet{Aigrain2016} and references therein. After applying \textit{k2sc}, persisting long-term trends -- if present -- were removed by fitting a low-degree polynomial. These may be physical but information regarding these trends is unprocurable in the limited \textit{K2} time base.

Generally the time span of each campaign is between 70 and 80~d. The frequency resolution of the extracted data is therefore $1/80$ \si{\day} $\sim 0.01$~d$^{-1}$ in the best case, which provides us with an upper limit on the frequency uncertainty when extracting independent pulsation mode frequencies from pre-whitening \citep{Aerts2010}. \textit{K2} Campaign~0 has two large gaps in the first 40~d, and generally only the last 30~d are usable. This results in a decrease in frequency resolution $1/30$ \si{\day} $\sim$ $0.03$~d$^{-1}$. For stars in Campaigns 9 and 11, where five of the eight of the stars in this study originate, a significant gap of $\sim$3~d occurs at BJD~24572694 in Campaign~9 and at BJD~24572847 in Campaign~11. This reduces the duty cycle. The different parts were extracted and detrended separately. For the frequency analysis they were stitched together again. All light curves are measured in long cadence at 29.5 minutes, resulting in a Nyquist frequency of $\nu_{\rm Ny} \approx 24.47 $~d$^{-1}$.

\subsection{Star sample}
To find new $\beta$~Cep stars with \textit{K2}, our sample comprises stars from \textit{Guest Observer Proposals} for massive star asteroseismology. The proposed targets were chosen based on their spectral types being either O or B in \texttt{SIMBAD}, and included stars in different stages of evolution. To separate the targets we viewed the periodograms of the reduced light curves and scanned for signs of coherent modes visually.
We filtered our sample, a total of 130 OB-type stars, for stars which show multi-periodic variability consistent with coherent pulsation modes, and excluded potential binaries as it is difficult to ascertain from which star the pulsations originate. The selected list of eight OB variables with dominant variability in the high-frequency regime observed by \textit{K2} is given in Table~\ref{table:Star_List}.

\subsection{Frequency analysis}\label{subsection: FA}
We perform \textit{iterative pre-whitening} as developed by \citet{Degroote2009a} and improved by \citet{Papics2012a} of the \textit{K2} light curves of stars in Table~1. At each step of the iterative pre-whitening we calculate the Lomb-Scargle (LS, \citealt{Lomb1976,Scargle1982}) periodogram of the light curve, followed by an identification of the frequency with the highest amplitude. This frequency value, along with previous frequencies, are used to perform a non-linear least-squares fit to the light curve using
\begin{equation}\label{eq:fitsignal}
x_{i}(t_{i}) = \sum_{j} A_{j} \sin\{2\pi[\nu_{j} (t_{i}-t_{0}) + \phi_{j}]\} + C,
\end{equation}
where $A_{j}$ and $\phi_{j}$ are the amplitude and phase, $\nu_{j}$ the frequency, and $t_{0}$ is the zero-point of a star's light chosen as the first observation. Here, $j = 1$, \dots, $n$ is the unknown number of pulsation frequencies and $i = 1$, \dots, $N$ with $N$ the number of data points in the observed light curve. After each iteration, the new multi-frequency model is then subtracted from the light curve, and a new iteration commences using the residuals.  This scheme is repeated until the signal-to-noise ratio (S/N) of a peak in the LS-periodogram, calculated in an interval of width 1~d$^{-1}$ centred around the frequency, is below five following \citet{Baran2015}. This criterion is based on simulations of the 90~d time base typical of \textit{K2}. Using this conservative criterion of S/N~$\geq 5$ coupled over a narrow frequency range where the dominant variability occurs, allows us to avoid over-interpretation of the data for a typical light curve in the sample, and prevents the introduction of spurious frequencies into the frequency list.

An important limitation for data sets of this time span is the frequency resolution, and careful consideration of the pre-whitened frequencies is recommended. \citet{Loumos1978} determined that two frequencies cannot be distinguished when their difference is less than 1.5 times the frequency resolution. We followed this criterion, i.e., when two significant frequencies are closer to each other than 1.5 times the resolution, the lowest amplitude peak is eliminated from the final frequency list. Combination frequencies of the form $n\nu_{i}+m\nu_{j}$, with $n,m$ $\in\mathbb{Z}$, may also be in the frequency spectra and have been detected for $\beta$~Cep stars \citep{Handler2004, Handler2006, Degroote2009a}. To search for combination frequencies we identify base frequencies as the dominant frequencies and scan for combinations of different orders with $n,m \in [-3,3]$ in a semi-automatic way. We use the frequency resolution (\si{1\per\Delta T}) as the matching criterion, and therefore list the values for each of the data sets.
It has been shown by \citet{Papics2012b} -- using simulated data and real examples -- that as the order of the combination increases, the number of frequencies identified as a combination increases and one should be careful not to misinterpret physical frequencies as combinations due to either a low frequency resolution or to high order combinations. 

By scanning for linear combination frequencies we are identifying mathematical combinations, irrespective of their origin. Due to the limited frequency resolution of \textit{K2} data we cannot infer a physical interpretation. If we list a frequency as a combination, it means that we identify it to be a mathematical match within the frequency resolution of the light curve. For a general discussion on the various physical origins of combination frequencies we refer to Chapter 6 of \citet{Bowman2017}.

\section{Results}\label{section: results}

In this section we present the results of the frequency extraction as described Section~\ref{subsection: FA}, in the form of light curves, periodograms and frequency lists. The errors on the frequencies and amplitudes are computed from the non-linear least-squares fitting described in Section~\ref{subsection: FA}. To correct for the correlated nature of over-sampled data, we use the correction factor as defined by \citet{Schwarzenberg-Czerny2003}, and calculated following \citet{Degroote2009a} and \citet{Papics2012a}. 
The frequency list of the $\beta$~Cep star EPIC~202060092 is given in Table~\ref{table:202060092_freq}, with the remaining frequency lists given in the Appendix~\ref{appendix: fl}. The errors on the frequencies and the amplitudes are given as the last digit, written in brackets after each value.
The last column in each table indicates if it has been identified as a combination or a harmonic.
In the caption of each table we include the standard deviation of the residuals, at the point where the next frequency drops below S/N = 5 in a 1~d$^{-1}$ window, as an indication of the noise level.

\begin{figure*}
    \centering
	\includegraphics[width=2\columnwidth, scale = 1]{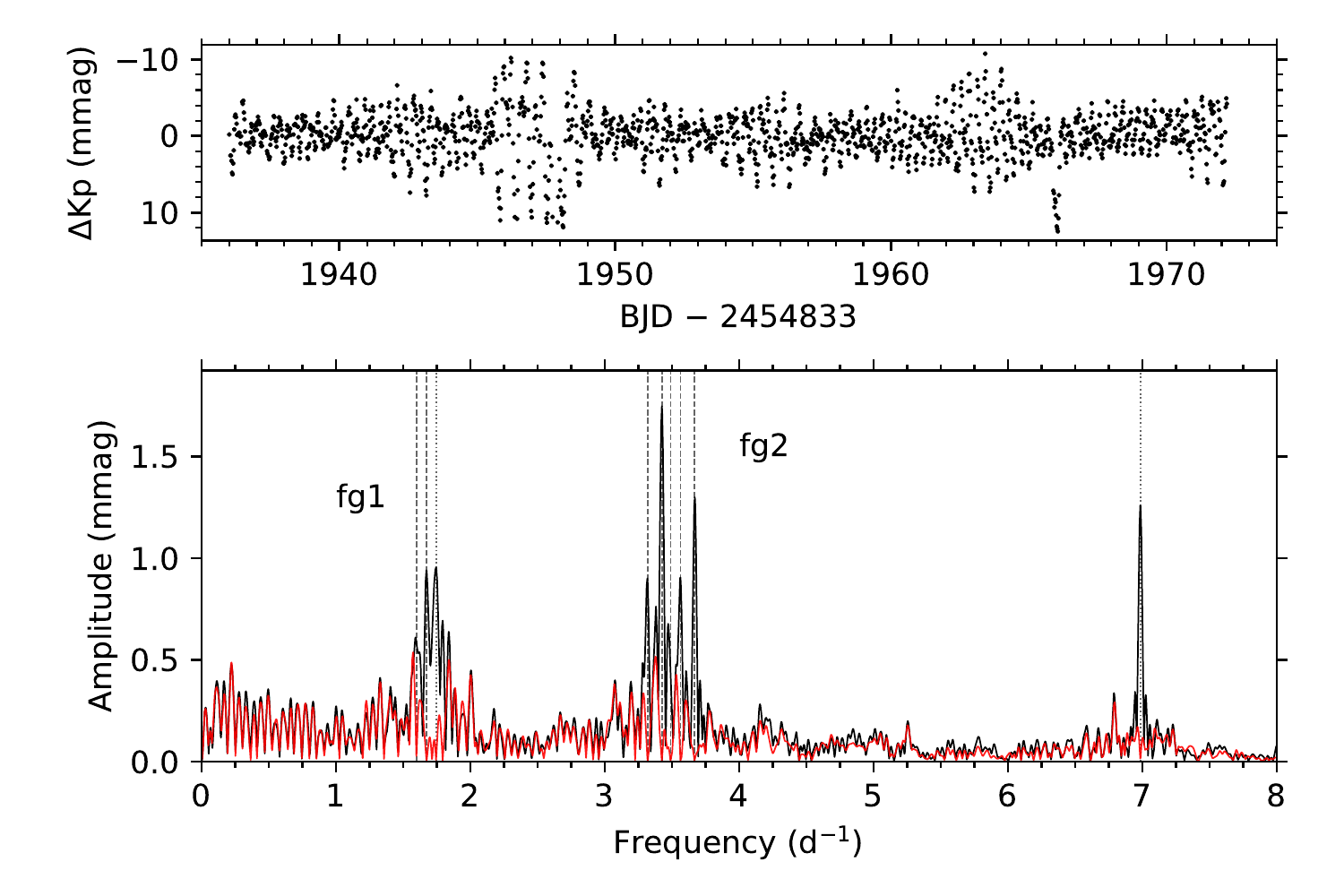}
    \caption{\textit{Top}: \textit{K2} light curve of EPIC~202060092 with the brightness variations in mmag. \textit{Bottom:} LS-periodograms and frequencies identified by pre-whitening. The initial periodogram is shown in black, while the periodogram of the residuals after pre-whitening is given in red. Dashed lines correspond to independent frequencies, the dotted lines represent low-order combinations/harmonics.}
    \label{fig:202060092_grid}
\end{figure*}

\subsection{EPIC~202060092 -- HD~256035}
In a spectroscopic study by \citet{Negueruela2004}, EPIC~202060092 was classified as O9V:p and it was suggested to be a spectroscopic binary because of its broad H$\alpha$ and HeI 6678 lines. The star was included in the sample of \citet{Buysschaert2015} in which it was confirmed to be a spectroscopic binary with at least one star in the system thought to be a  $\beta$~Cep star based on the frequency spectrum. The dominant frequency is found to be $\nu = 3.424(5)$~d$^{-1}$ and other mode frequencies occur in the range $0.11 \leq \nu \leq 6.99$~d$^{-1}$. 

Because of the two major gaps in the first 40~d of the \textit{K2} photometry, only the data points after 1936.0~(BJD-2454833.0) were extracted and run through \textit{k2sc}. This yielded a total of 36.19~d and 1679 data points resulting in a duty cycle of $\sim 97$~per~cent. The frequency resolution is 0.028~d$^{-1}$. The detrended light curve is given in the top panel of Fig.~\ref{fig:202060092_grid} and the LS-periodogram is shown in the bottom panel. Dashed lines indicate independent frequencies, and dotted lines identify combination frequencies. 

Iterative pre-whitening reveals nine significant frequencies with amplitudes ranging between 0.6(1)~mmag and 1.9(2)~mmag which is the lowest dominant peak in our sample. Frequency groups are seen between $1.5 \leq \nu \leq 2$~d$^{-1}$ (labelled as frequency group 1, fg1), and between $3 \leq \nu \leq 4$~d$^{-1}$ (labelled as frequency group 2, fg2). Some frequencies in one group can be found as combinations of the other group, and vice versa, as was already noted by \citet{Buysschaert2015} (see their Figure 1 and Table 2). Due to the poor frequency resolution we cannot determine whether these combinations are real or coincidental. This phenomenon is a general feature of the analysed stars and is due to the limited time span of the \textit{K2} data. 

Despite the additional detrending with \textit{k2sc}, the light curve does not yield more identified frequencies than found by \citet{Buysschaert2015}; our list consists of nine. It can be seen from the red curve in the bottom panel of Fig.~\ref{fig:202060092_grid} that variability is still present in the residuals. However, pre-whitening beyond our stop criterion leads to unresolved frequencies and is therefore omitted.
It should also be noted that the frequency lists do not match with \citet{Buysschaert2015}, they identify additional low frequencies < 1~d$^{-1}$ (as harmonics) and one at $1.32(1)$~d$^{-1}$. Frequency $\nu_{8} = 1.599(3)$~d$^{-1}$ in our list is not detected by them. The differences can largely be explained by the extraction and detrending procedure; \citet{Buysschaert2015} did not use \textit{k2sc}. This highlights the importance and impact of detrending on extracted pulsation modes in variable stars and any subsequent frequency analysis. Clearly, data covering a longer time span are needed to fully resolve and characterise the frequency groups, and confirm EPIC~202060092 as a g- or p-mode dominated star.

\begin{table}
\caption{Frequency list for EPIC~202060092 extracted by pre-whitening. The term between brackets in the ID column denotes which group the particular frequency belongs to.  The standard deviation of the residuals is 0.068 mmag.} \label{table:202060092_freq}      
\centering          
\begin{tabular}{c c c c c}  
\hline

ID &	Frequency 	& Amplitude	&SN & Notes \\
 & [\si{\per\day}] & [mmag] & & \\
\hline

$\nu_{1}$ (fg2)&	3.424(2)&	1.9(2)&	9.8&\\
$\nu_{2}$ (fg2)&	3.667(2)&	1.3(2)&	8.4&\\
$\nu_{3}$ &	6.985(2)&	1.2(2)&	18.2& $\nu_{2}+\nu_{5}$, 2$\nu_{9}$, 4$\nu_{6}$\\
$\nu_{4}$ (fg1)&	1.674(2)&	1.2(2)&	5.8& \\
$\nu_{5}$ (fg2)&	3.319(3)&	1.0(2)&	6.2& 2$\nu_{4}$\\
$\nu_{6}$ (fg1)&	1.745(3)&	1.0(2)&	6.0& \\
$\nu_{7}$ (fg2)&	3.563(3)&	0.9(2)&	6.5&\\
$\nu_{8}$ (fg1)&	1.599(3)&	0.7(1)&	5.0& \\
$\nu_{9}$ (fg2)&	3.489(3)&	0.7(1)&	5.4&\\

\hline
\end{tabular}
\end{table}

\begin{figure*}
	\includegraphics[width=2\columnwidth, scale = 1]{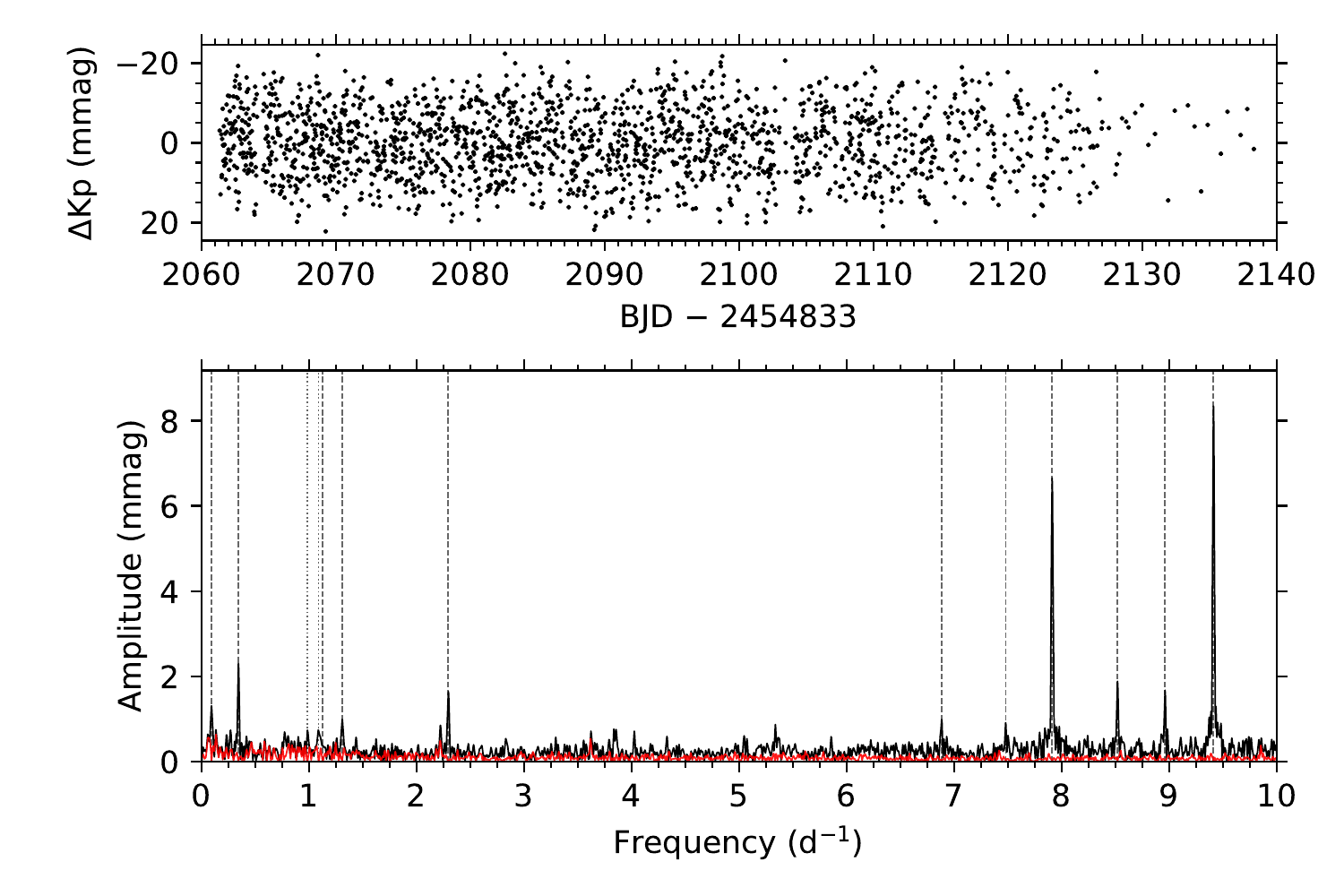}
    \caption{\textit{Top}: \textit{K2} light curve of EPIC~202691120 with the brightness variations in mmag. \textit{Bottom:} LS-periodograms and frequencies identified by pre-whitening. The linestyles are the same as in Fig.~\ref{fig:202060092_grid}. No significant frequencies beyond 10 d$^{-1}$ are detected.}
    \label{fig:202691120_grid}
\end{figure*}

\subsection{EPIC~202691120 -- CD-28 12286}\label{sec:202691120}
This star was observed by \textit{K2} in Campaign~2 and classified by \citet{Drilling1995} as OB$^{-}$ in their galactic OB survey. Stars that showed nearly continuous spectra were assigned spectral OB, with OB$^{+}$, OB, OB$^{-}$ indicating increasing Balmer absorption-line strength. This is based on the original photographic plate classification scheme by \citet{Nassau1960}, who segregated OB stars based on hydrogen-line strength and used this to classify any star earlier than B6. The authors do note that late B-type  and early A-type supergiants may be misclassified as OB, as their hydrogen line imitates that of the OB stars. The presence of the Ca-II line is used to segregate the A stars, such that main sequence A stars are unlikely to be classified as OB.
A total of 1904 \textit{K2} data points were extracted, over 76.94~d, which implies a duty-cycle of only $\sim51$~per~cent.  This is a result of a large number of poor quality flags near the end of the time series, resulting in their removal by the \textit{k2sc} module. The detrended light curve is given in the upper panel of Fig.~\ref{fig:202691120_grid}. The LS-periodogram, shown in the bottom panel of the same figure, indicates frequencies between 0 and 10~d$^{-1}$.

A total of thirteen frequencies were extracted by the pre-whitening, which have amplitudes between 0.7(1) and 8.3(3)~mmag. Two of the lower frequencies are combinations, leaving five independent ones below 2~d$^{-1}$ and six above 6~d$^{-1}$.  Therefore we interpret the independent low frequencies as low amplitude g-modes. The higher frequencies are coherent oscillations. If we take the differences between successive frequencies in the high frequency regime we retrieve three splittings of $\sim 0.44$~d$^{-1}$ , two between the three highest frequencies up to $\nu_{1}$ and one between $\nu_{2}$ and a lower amplitude frequency. In addition, two splittings of $\sim0.60$~d$^{-1}$ become apparent, one next to $\nu_{2}$ and one between two lower amplitude frequencies. These equidistant frequencies might be due to rotational splitting. Of course we cannot unambiguously claim that these frequencies belong to the same multiplets, see for example the identification of a previously claimed multiplet to three different modes in the $\beta$~Cep star 12 Lacertae \citep{Handler2006}. Other methods such as multi-passband photometry are needed to confidently assign the geometries of these modes.
The frequency list is given in Table~\ref{table:202691120_freq}, demonstrating that EPIC~202691120 is predominantly a p-mode pulsator. The presence of low frequencies that cannot be identified as combination frequencies of the p-modes makes us classify the star as a hybrid $\beta$~Cep/SPB.

\begin{figure*}
	\includegraphics[width=2\columnwidth, scale = 1]{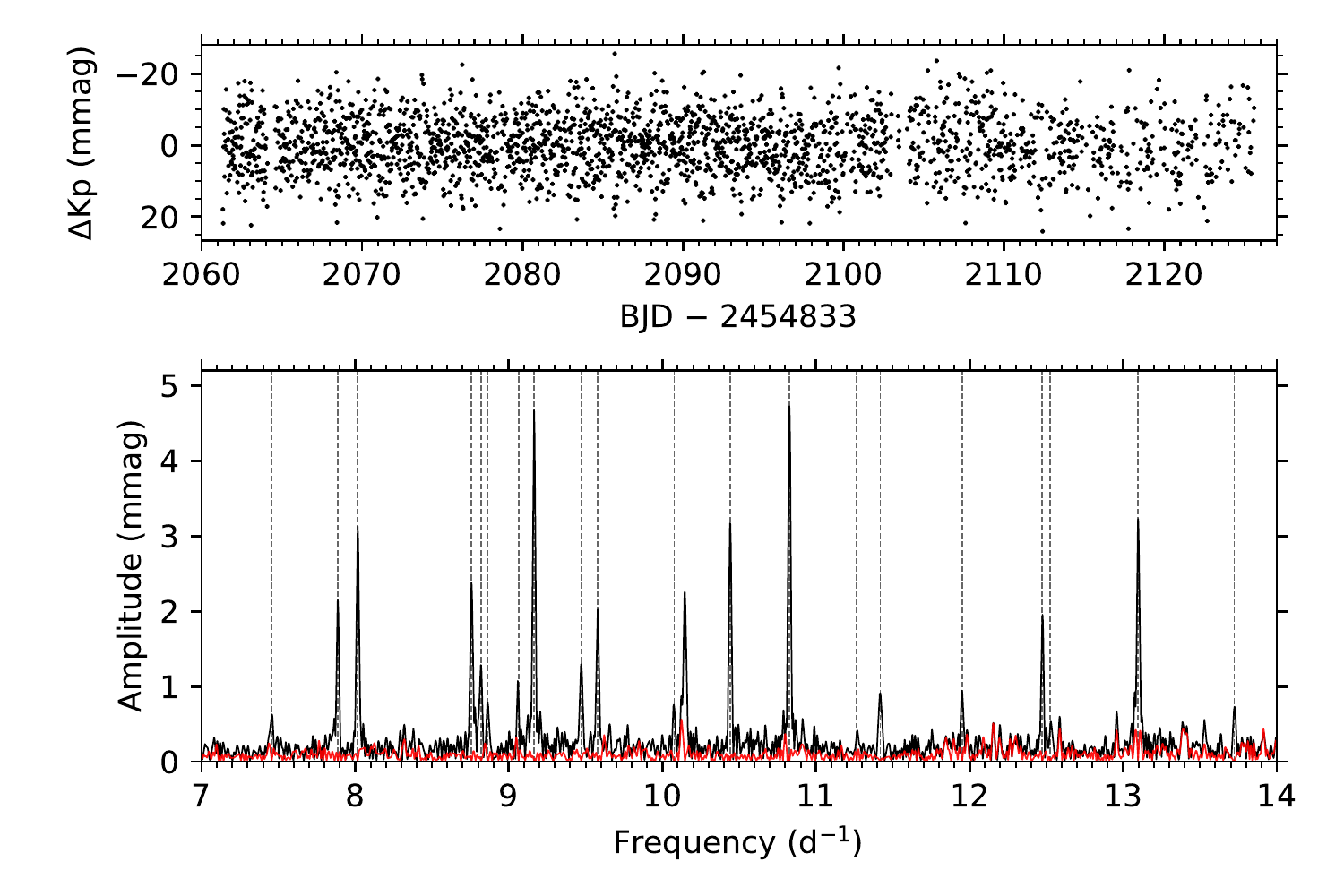}
    \caption{\textit{Top}: \textit{K2} light curve of the newly discovered $\beta$~Cep star EPIC~202929357 with the brightness variations in mmag. \textit{Bottom:} LS-periodograms and frequencies identified by pre-whitening. The linestyles are the same as in Fig.~\ref{fig:202060092_grid}.}
    \label{fig:202929357_grid}
\end{figure*}

\subsection{EPIC~202929357 -- CD-27 10876}\label{sec:202929357}
This star was classified by \citet{Drilling1995} as having a spectral type of OB$^{-}$. The \textit{K2} Campaign~2 light curve and LS-periodogram are shown in Fig.~\ref{fig:202929357_grid}, in the upper and bottom panels respectively. The detrended light curve contains 2302 data points over 64.25~d. This yielded a duty cycle of $\sim75$~per~cent. The frequency resolution is 0.015~d$^{-1}$. 

Pre-whitening leads to a total of 23 frequencies. Nearly all the frequencies detected in EPIC~202929357 are detected in the high frequency regime > 7~d$^{-1}$. Only two lower frequencies, $\nu_{15} = 4.0458(9)$~d$^{-1}$ and $\nu_{16} = 0.051(1)$~d$^{-1}$ are identified. The amplitudes range between 0.6(1) and 4.7(2)~mmag.  In the periodogram in Fig.~\ref{fig:202929357_grid} we only show the high-frequency region for good visibility. No low-order combinations are detected so we infer that all identified frequencies are due to independent modes. The frequencies were scanned for similar spacings in order to detect possible rotational splitting, but none could be identified unambiguously.  

Similarly to EPIC~202691120, EPIC~202929357 has high-frequency p-modes indicative of a young massive $\beta$~Cep star.
Moreover, only two frequencies are detected below 5~d$^{-1}$ such that the star is predominantly a p-mode pulsator. A longer time series, high-resolution spectroscopy, or multi-colour photometry, might aid in mode identification. Nonetheless, the coherent character of the high-frequency modes, together with the spectral type OB, allows us to classify this star as a new $\beta$~Cep variable. 

Despite the rough OB classification and lack of spectroscopy we do need to address whether the presence of p-mode pulsations is enough to classify the Campaign~2 stars EPIC~202691120 and EPIC~202929357 as $\beta$~Cep. Lower mass stars (A2 -- F2), in the $\delta$~Scuti instability region show variability on similar time-scales \citep{Rodriguez2001} and \texttt{SIMBAD} gives estimates between 7500-9300~K for EPIC~202691120 and EPIC~202929357. However, these values are from \citet{Huber2016}, who use synthetic star populations sampled from distributions lacking massive and pre-main sequence stars. The authors encourage users of the EPIC catalogue to use other observational information instead -- such as colours -- to preliminary characterise these types of stars \citep{Huber2016} in absence of derived spectroscopic parameters.
We therefore verify the $\beta$~Cep classification by using Gaia-DR2 photometry and by placing the stars in a colour-magnitude diagram, see Section \ref{section: discussion}, confirming the \texttt{SIMBAD} classification as OB.

\begin{figure*}
	\includegraphics[width=2\columnwidth, scale = 1]{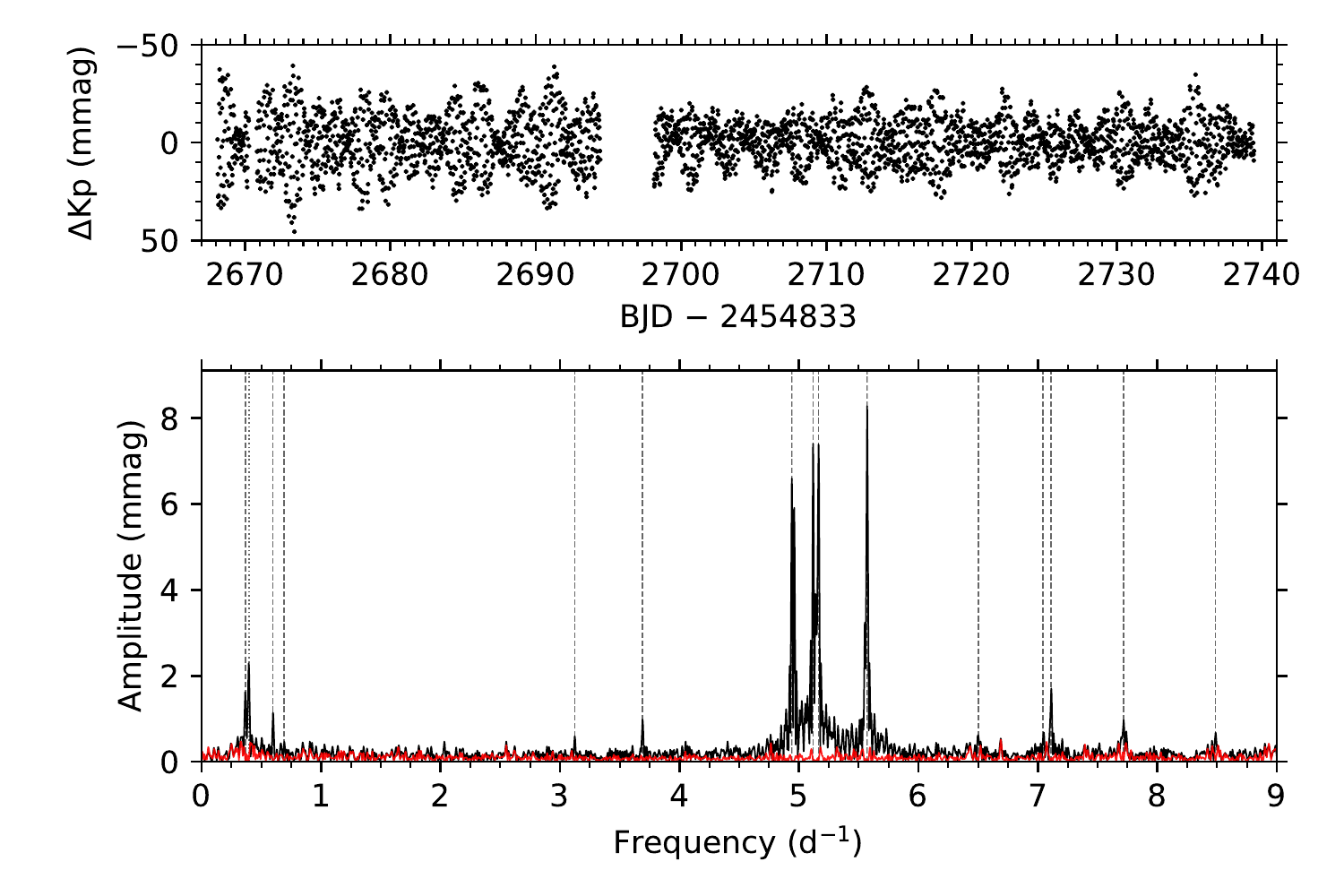}
    \caption{\textit{Top}: \textit{K2} light curve of EPIC~223832867 with the brightness variations in mmag. \textit{Bottom:} LS-periodograms and frequencies identified by pre-whitening. The linestyles are the same as in Fig.~\ref{fig:202060092_grid}.}
    \label{fig:223832867_grid}
\end{figure*}
 
\subsection{EPIC~223832867 -- HD~164741}\label{sec:223832867}
The object is classified as a B2Ib/II star \citep{Houk1988} and was identified as a $\beta$~Cep by \citet{Pigulski2008} with a dominant frequency of 5.119632(19)~d$^{-1}$. They report four additional independent modes, with a detection threshold of 3.9~mmag, using 474 ground-based observations covering $2847$~d. EPIC~223832867 was also studied by \citet{Hohle2010}, who calculated the masses and luminosities of OB stars detected in the 2MASS and Hipparcos surveys using multicolour photometry, spectral types, extinctions and revised Hipparcos distances. The estimated mass of EPIC~223832867, $6.3 \pm 0.4$~M$_{\odot}$, is typically lower than expected for $\beta$~Cep pulsators but many factors can influence the exact boundaries of the theoretical $\beta$~Cep instability region (see \citealt{Miglio2007b}).  The bolometric luminosity and the effective temperature have also been estimated to be log($L/L_{\sun}) = 3.12 \pm 0.03 $ and log~$T_\text{eff} = 4.28^{+0.04}_{-0.05}$~K. The star is classified as a Be star \citep{Jaschek1982, Steele1999, Barnsley2013}, based on the H$\alpha$ emission. The projected rotational velocity of the object was determined to be $v\,\sin\,i = 77$~km~s$^{-1}$ \citep{Steele1999}, which is moderate so it is likely seen at low-inclination. 

Stitching the two sub-campaign \textit{K2} light curves of EPIC~223832867  yields 3055 data points over 71.34~d, resulting in a duty cycle of $\sim89$~per~cent. The frequency resolution is 0.014~d$^{-1}$. The light curve is shown in the upper panel of Fig.~\ref{fig:223832867_grid} and shows both short-term variations and a long-term beating pattern. An amplitude difference between the sub-campaigns is noticeable. We do not observe this in the other Campaign~9 star, EPIC~227552090 (Fig.~\ref{fig:227552090_grid}), and therefore argue that it is unlikely instrumental.  
The LS-periodogram in the bottom panel of Fig.~\ref{fig:223832867_grid} shows low ($ \leq 1$~d$^{-1}$) and high ($4 \leq \nu \leq 8$~d$^{-1}$) frequencies. A total of fifteen are identified from pre-whitening. The amplitudes range between 0.5(1)~mmag and 8.4(5)~mmag.  The first four high amplitude frequencies all appear as a doublet structure in the periodogram. The close proximity of the frequencies in these doublets leads to the \citet{Loumos1978} criterion being satisfied (< 1.5/$\Delta$T, with $\Delta$T the time base of the light curve) for three of the eight members. We do not consider them in the frequency list as they are formally unresolved from each other given this criterion, although they are likely real. Only one doublet is sufficiently separated to be kept as two frequencies in the frequency list. The light curve in Fig.~\ref{fig:223832867_grid} exhibits beating patterns that are likely caused by unresolved close-frequency pulsations at both high and low frequency (see e.g. \citealt{Bowman2016}), combined with rotation. The low frequency peak with the highest amplitude is identified as the combination frequency, $\nu_{5} \approx \nu_{1} - \nu_{2} = 0.3949(6)$~d$^{-1}$.

Further data, spanning a longer time span are necessary to fully characterise the beating patterns produced by the close-frequency pulsation modes in EPIC~223832867. The frequency list is given in Table~\ref{table:223832867_freq} and is consistent with the earlier results by \citet{Pigulski2008}. We denote the frequencies identified by them in the last column in Table~\ref{table:223832867_freq}.

\subsection{EPIC~227552090 -- HD~169173}\label{section: 227552090res}

\begin{figure*}
	\includegraphics[width=2\columnwidth, scale = 1]{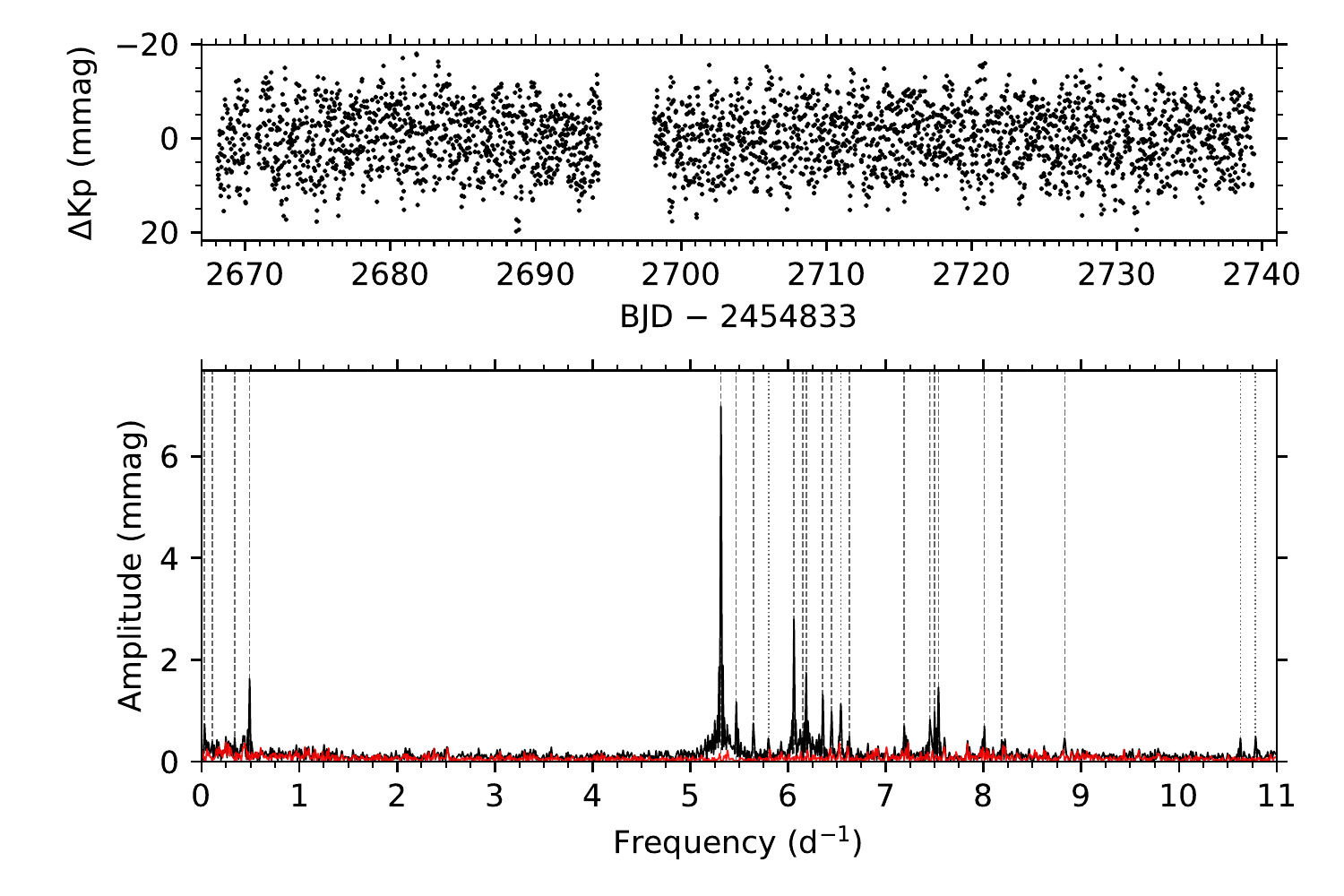}
    \caption{\textit{Top}: \textit{K2} light curve of EPIC~227552090 with the brightness variations in mmag. \textit{Bottom:} LS-periodograms and frequencies identified by pre-whitening. The linestyles are the same as in Fig.~\ref{fig:202060092_grid}.}
    \label{fig:227552090_grid}
\end{figure*}

The star has spectral type B3Ib \citep{Houk1988}. \citet{Pigulski2008} identified this star as a $\beta$~Cep star, finding only one frequency ($\nu_{1}= 5.31303(3)$~d$^{-1}$) using 789 data points of ASAS-3 photometry.

\begin{figure}
	\includegraphics[width=1\columnwidth, scale = 1]{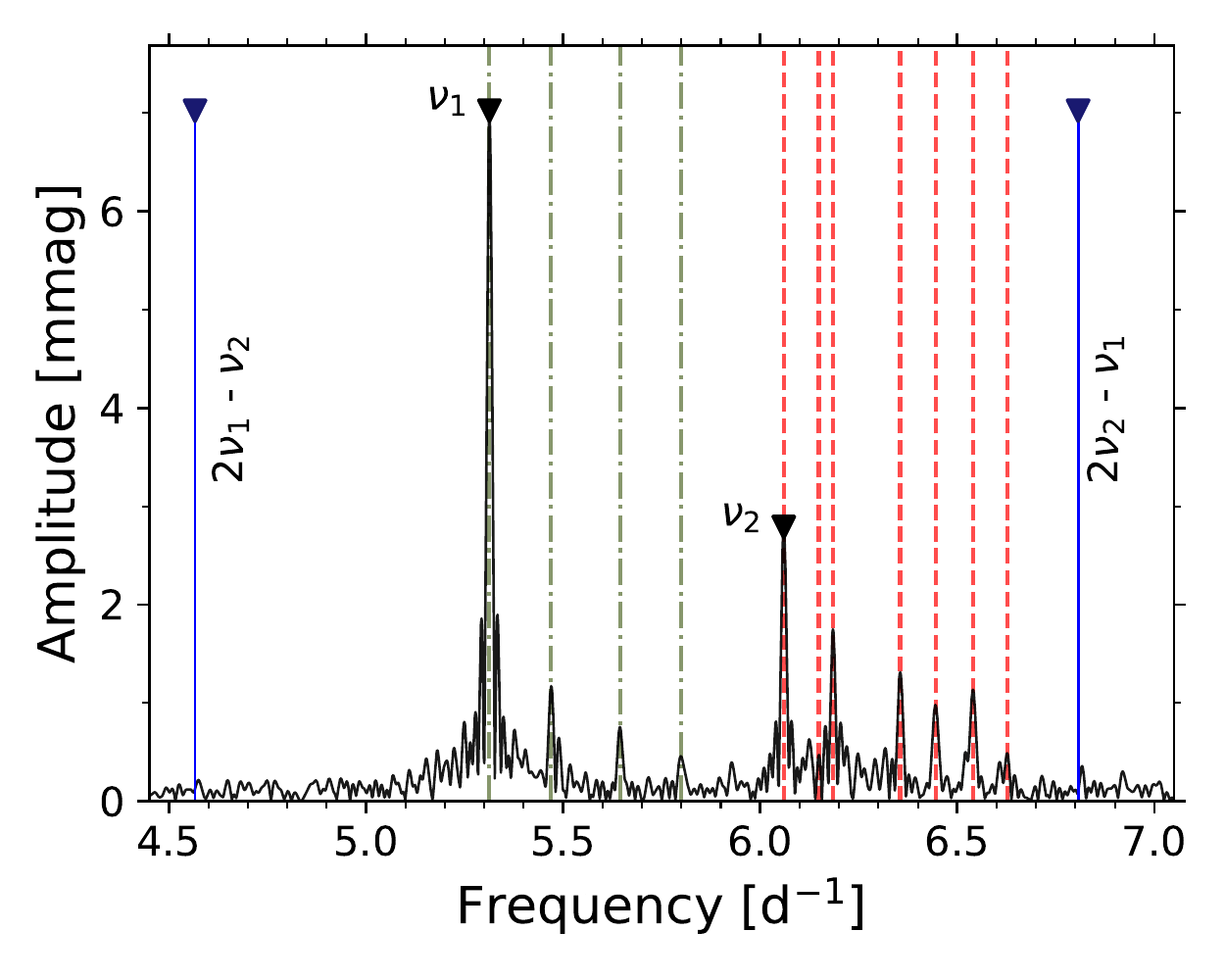}
    \caption{Zoom-in of the region where frequencies equally spaced in frequency are detected in EPIC~227552090. The dashed/dash-dotted lines are resolved frequencies, and the blue lines depict locations of generated combinations of $\nu_{1}$ and $\nu_{2}$ (marked with a black triangle) that are possible within this frequency range. The green dash-dotted and the red dashed lines indicate frequencies referred to as the first series and second series respectively.}
    \label{fig:227552090_combos}
\end{figure}

The \textit{K2} light curve EPIC~227552090 is given in the upper panel of Fig.~\ref{fig:227552090_grid} and consists of 2964 data points over 71.34~d. The duty cycle is therefore $\sim86$~per~cent and the frequency resolution is 0.014~d$^{-1}$. The LS-periodogram, shown in the lower panel of Fig.~\ref{fig:227552090_grid}, reveals a dominant frequency of $\nu_{1} = 5.3131(2)$~d$^{-1}$ confirming the result by \citet{Pigulski2008}. We detect a total of 24 frequencies with amplitudes ranging between 0.39(8) and 7.0(2)~mmag, of which three have been identified as possible combination frequencies.  We identify two separate frequency series, visible in Fig.~\ref{fig:227552090_combos}. The first series is marked by the green dashdot lines and the second by the red dashed lines. The first series starts at $\nu_{1}= 5.3131(2)$~d$^{-1}$, the dominant peak in the LS-periodogram, and the second series at the second dominant peak $\nu_{2}=6.0599(4)$~d$^{-1}$.  The spacings are not completely regular, in the first series they average around $\sim$0.16~d$^{-1}$. In the second series the spacings range from $\sim$0.08~d$^{-1}$ to $\sim$0.09~d$^{-1}$, all within the frequency errors, if we ignore the first three peaks of the second series in Fig.~\ref{fig:227552090_combos}. On the other hand, this second series could very well also be a rotational quintuplet with a missing member, unrelated to the first series.   

Furthermore there is a decrease in amplitude for increasing frequency leading to an asymmetric frequency pattern (see Fig.~\ref{fig:227552090_combos} for a zoom-in of the region where the spacings occur). 
In each series one member is found as a combination of the highest amplitude peak in that series, that is $\nu_{1}$ or $\nu_{2}$, with the highest amplitude peak in the low frequency regime $\nu_{4}=0.4907(6)$~d$^{-1}$ (i.e. $\nu_{16} = \nu_{1}+\nu_{4}$ and $ \nu_{7} = \nu_{2}+\nu_{4}$). These are given in the last column of Table~\ref{table:227552090_freq} and shown as the dotted lines in Fig.~\ref{fig:227552090_grid}.

Approximately equally-spaced p-mode frequencies have been detected in some massive stars, for example by \citet{Degroote2010b} for HD~46149 where a spacing of $\Delta \nu=0.48 \pm 0.02$ \si{\per\day} was uncovered for short lifetime modes with frequencies between $3.0 < \nu < 7.5$~d$^{-1}$. Similar oscillations have also been claimed by \citet{Belkacem2009} to occur in V1449~Aql, a star previously classified as a large amplitude $\beta$~Cep pulsator. \citet{Degroote2009a} proposed non-linear resonant mode excitation as the origin of the frequency spacing pattern instead. They showed that many frequencies in V1449~Aql appear as combinations of the dominant modes and are phase-locked ($\phi_{z} = \phi_{x} \pm \phi_{y}$, where $x,y$ indicate the parent frequencies and $z$ the combination), which is not expected for stochastically excited modes. The regular frequency spacing in V1449~Aql was fully reconstructed from combinations of the high amplitude modes \citep{Degroote2009a}. 

Also EPIC~227552090 has a dominant mode, twice as large as the next frequency in terms of amplitude. This raises the question whether non-linear resonant mode excitation may also be at work here. Indeed, asymmetric patterns are certainly possible through non-linear resonant mode coupling, as seen in white dwarfs \citep{Buchler1995}. To verify this we assumed that the two dominant modes in EPIC~227552090, $\nu_{1}=5.3131(2)$~d$^{-1}$ and $\nu_{2}= 6.0599(4)$~d$^{-1}$, are the parent frequencies and generated combinations up to the fifth order. As shown in Fig.~\ref{fig:227552090_combos}, none of the generated combination frequencies fall within the frequency resolution of the resolved frequencies. We therefore conclude that non-linear resonant mode excitation is not the origin of the regular frequency spacing in EPIC~227552090. 

\subsection{EPIC~233986359 -- HD~156491}

\begin{figure*}
	\includegraphics[width=2\columnwidth, scale = 1]{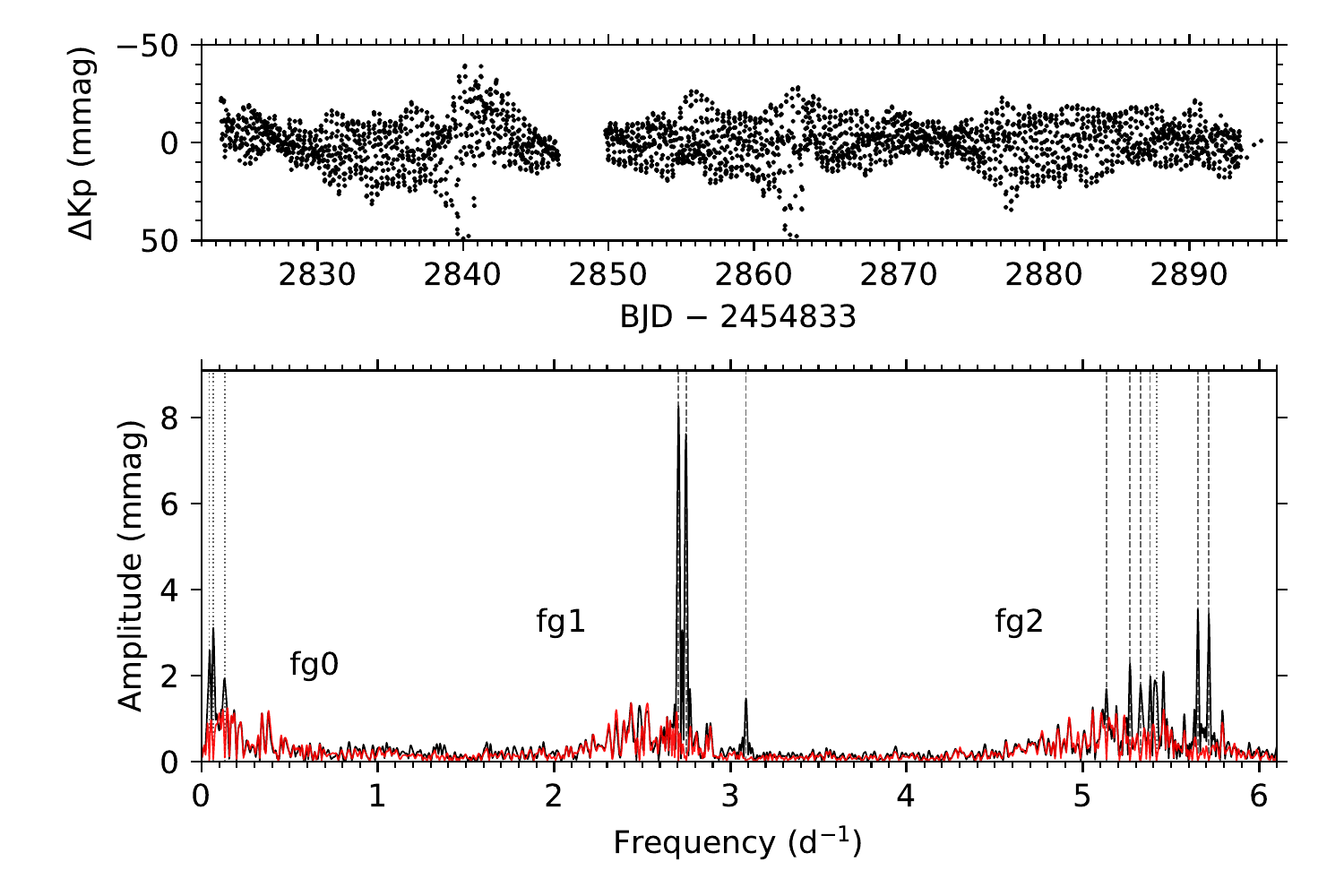}
    \caption{\textit{Top}: \textit{K2} light curve of EPIC~233986359 with the brightness variations in mmag. \textit{Bottom:} LS-periodograms and frequencies identified by pre-whitening. The linestyles are the same as in Fig.~\ref{fig:202060092_grid}.}
    \label{fig:233986359_grid}
\end{figure*}

This star is spectral type B3III \citep{Houk1988}. It shows emission in H $\alpha$ \citep{MacConnel981} and is therefore classified as a spectroscopic Be star. The \textit{K2} light curve of EPIC~233986359 is shown in the upper panel of Fig.~\ref{fig:233986359_grid}. A total of 2957 data points over 71.59~d yields a duty cycle of $\sim86$~per~cent and a frequency resolution of 0.014~d$^{-1}$. The LS-periodogram, shown in the lower panel Fig.~\ref{fig:233986359_grid}, reveals the presence of distinct frequency groups labelled as fg0, fg1 and fg2. We extract a total of thirteen frequencies with amplitudes between 1.4(3)~mmag and 8.2(5)~mmag. The frequency list is presented in Table~\ref{table:233986359_freq}.

The light curve exhibits complex beating patterns caused by the close spacing of frequencies in the two frequency groups, both shown in Fig.~\ref{fig:233986359_grid}.
The close frequency spacing pattern implies a high local noise estimate. The stop-criterion is satisfied rapidly, such that the number of resolved frequencies is limited. We cannot reconstruct all frequencies in one group as combination frequencies from another group. Only the harmonic of the dominant frequency (in terms of amplitude) is found in fg2, $\nu_{8} = 2 \nu_{1}$. The low frequencies in fg0 on the other hand can all be explained as differences of the frequencies fg1 suggesting that there may also be independent mode frequencies above 4~d$^{-1}$. The detected combinations are given in a separate column in Table~\ref{table:233986359_freq}. Fundamental parameters from spectroscopy are needed to model this interesting Be star.

\subsection{EPIC~235094159 -- LS 3978}

\begin{figure*}
	\includegraphics[width=2\columnwidth, scale = 1]{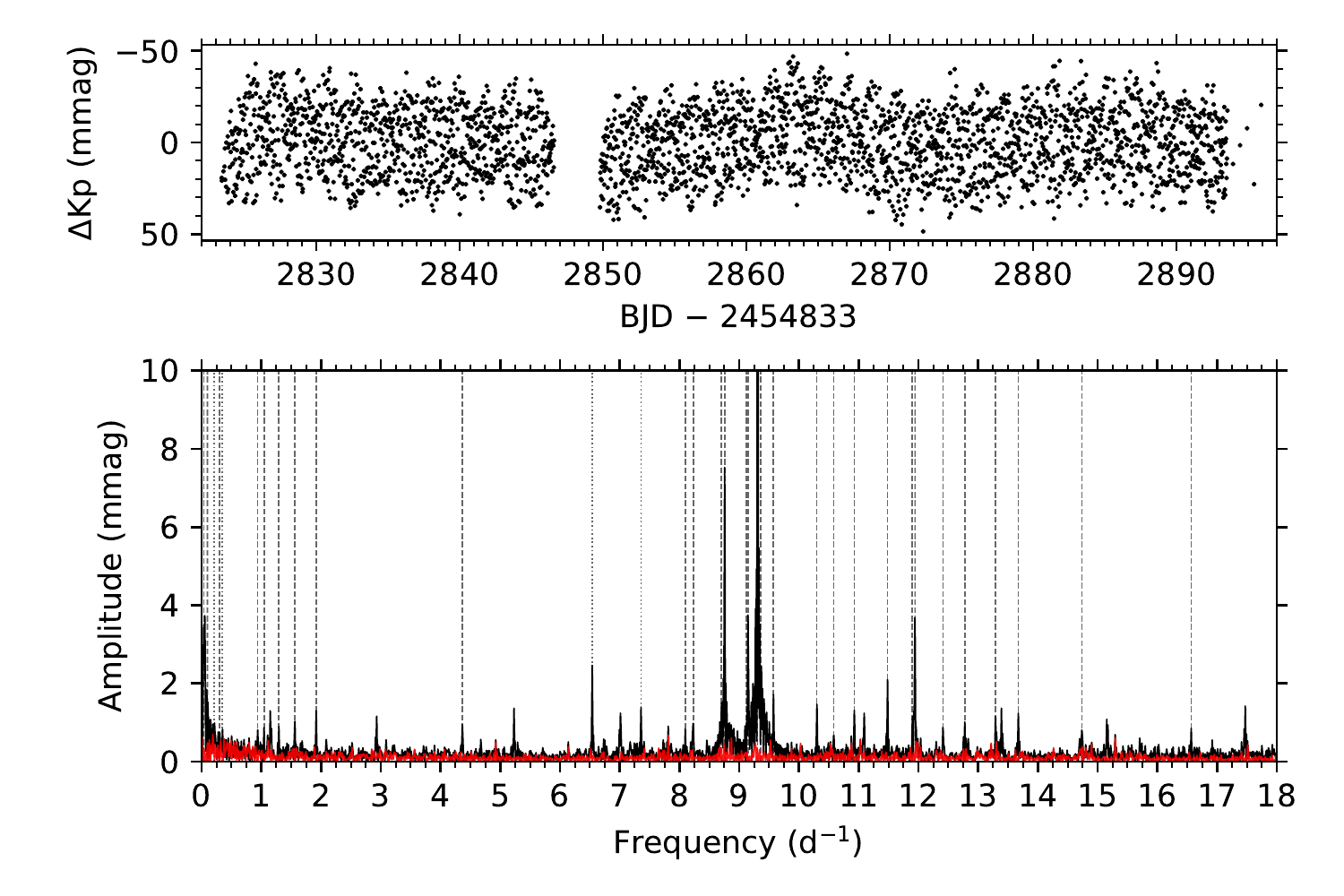}
    \caption{\textit{Top}: \textit{K2} light curve of EPIC~235094159 with the brightness variations in mmag. \textit{Bottom:} LS-periodograms and frequencies identified by pre-whitening. The amplitude is cut off at 10~mmag, the dominant peak has an amplitude of 21.3(3)~mmag.
    The linestyles are the same as in Fig.~\ref{fig:202060092_grid}.}
    \label{fig:235094159_grid}
\end{figure*}

This star is of spectral type B2III \citep{Vijapurkar1993}.
The \textit{K2} light curve is shown in the upper panel of Fig.~\ref{fig:235094159_grid} and consists of 3012 data points over 72.51~d, resulting in a duty cycle of $\sim87$~per~cent and a frequency resolution of 0.014~d$^{-1}$. The amplitudes of the oscillations are large compared to other stars in our sample. This is advantageous as this is the faintest object (in terms of V magnitude, see Table~\ref{table:Star_List}), resulting in a higher noise level than for the other stars. Nonetheless, the LS-periodogram shown in the bottom panel of Fig.~\ref{fig:235094159_grid} reveals many independent mode frequencies. 

Pre-whitening reveals a total of 35 frequencies with amplitudes between 0.7(1) and 21.3(3)~mmag. For visibility the vertical axis in the bottom panel of Fig.~\ref{fig:235094159_grid} was cut to 10~mmag. We scanned for combinations but no obvious candidates were detected. The frequency with the highest amplitude below 1~d$^{-1}$, that is $\nu_{3} = 0.0381(4)$ i.e. a period of $\sim26.25$~d, shows multiple harmonics. In fact, many low frequencies <1~d$^{-1}$ can be explained as harmonics of $\nu_{3}$ up to 9$\nu_{3}$, but 4$\nu_{3}$, 5$\nu_{3}$, 6$\nu_{3}$ and 7$\nu_{3}$ are missing. We cannot identify further frequencies below $4$~d$^{-1}$, besides the harmonic series, as combinations in the high frequency regime. 

The frequency list is given in Table~\ref{table:235094159_freq}. The detected harmonics are indicated in the last column of the same table and suggest that this is a pulsating B star in a binary with ellipsoidal variability or a single B star with  rotational modulation, in addition to high frequency modes. Unfortunately information about the spherical degree $\ell$ and radial orders $n$ cannot be retrieved so the only current source for seismic input is the frequency list presented here.

\begin{figure*}
	\includegraphics[width=2\columnwidth, scale = 1]{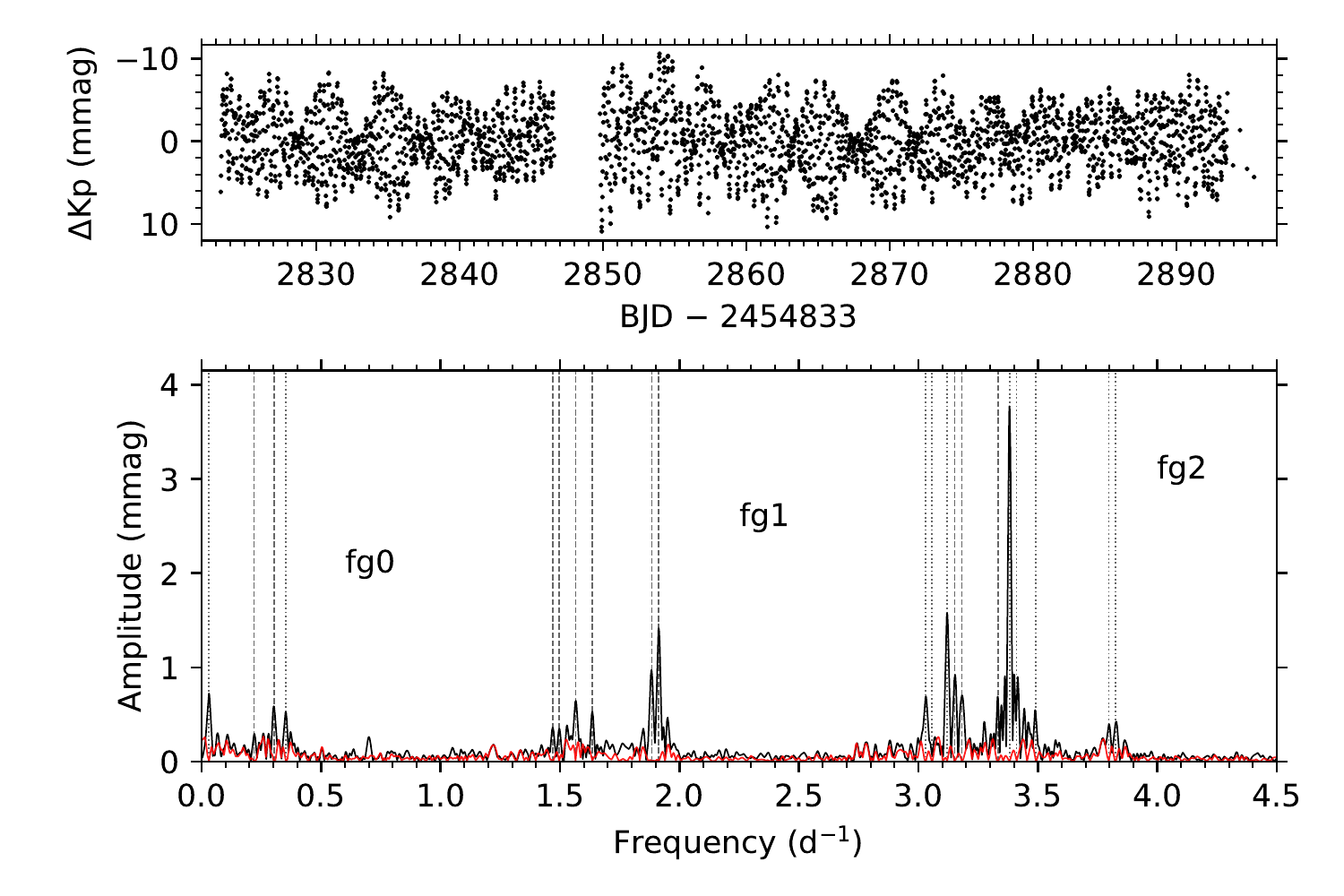}
    \caption{\textit{Top}: \textit{K2} light curve of EPIC~235151005 with the brightness variations in mmag. \textit{Bottom:} LS-periodograms and frequencies identified by pre-whitening. The linestyles are the same as in Fig.~\ref{fig:202060092_grid}. }
    \label{fig:235151005_grid}
\end{figure*}

\subsection{EPIC~235151005 -- HD~157548}\label{sec:235151005}
This star has spectral type B5III according to \citet{Houk1988}. The \textit{K2} light curve of EPIC~235151005, and its corresponding LS-periodogram are shown in Fig.~\ref{fig:235151005_grid}. The light curve contains 2971 data points over 72.08~d resulting in a duty cycle of $\sim86$~per~cent and a frequency resolution of 0.014~d$^{-1}$. A total of 21 significant frequencies are detected, with amplitudes between 0.29(7)~mmag and 3.7(2)~mmag. In the LS-periodogram we distinguish three distinct groups of low frequencies (< 4~d$^{-1}$), in order of increasing frequency values. This appearance of frequencies in closely-spaced groups is similar to EPIC~202060092 and EPIC~233986359. We cannot explain all frequencies in fg1 as combinations of frequencies in fg2, or vice versa.

A category of pulsating stars that show frequencies occurring in distinct groupings are Be type variables, broadly defined as B-type stars showing emission in H$\alpha$. \citet{Saio2007}, \citet{Neiner2009} and \citet{Balona2011} noted several instances of frequency groupings in Be stars in a large study of B-type variables observed with the \textit{MOST, CoRoT} and \textit{Kepler} space missions respectively. These stars are characterised as fast rotators that may feed  mass to circumstellar disks in episodic mass ejections \citep{Porter2003}. The irregular outbursts, in the case of HD~49330, have been hypothesised by \citet{Huat2009} as occurring when the close interacting frequencies come into phase with each other. The Be star studied by \citet{Kurtz2015}, KIC~11971405, has an amplitude spectrum with striking resemblance to the amplitude spectrum for EPIC~235151005 (Fig.~\ref{fig:235151005_grid}). Similar frequency patterns are also seen in the B5IVe star HD~181231 by \citet{Neiner2009}, albeit at lower frequencies. In these studies, the authors were able to explain most frequencies in other groups as combinations from a limited set of parent frequencies, identified as g-modes.

In the stars showing frequency group structure (EPIC~202060092, EPIC~233986359, EPIC~235151005) we were not able to interpet all frequencies from a set of parent frequencies. We therefore argue in favour of a hybrid pulsation character as seen in \textit{CoRoT} B2V star HD~170580 \citep{Aerts2019b} for which the authors similarly were not able to explain all high frequencies as combinations. In this star, also distinct frequency groups were detected, between $1.4 \leq \nu \leq 2$~d$^{-1}$ and $3.1 \leq \nu \leq 3.9$~d$^{-1}$, remarkably similar to what we find for EPIC~235151005.  For that very slowly rotating star the frequencies were identified as modes with radial orders $n$~$\in$[-7,2]. Future spectroscopy will reveal whether these three \textit{K2} pulsators are fast or slow rotators.

\section{Discussion} \label{section: discussion}

\begin{figure*}
	\includegraphics[width=2\columnwidth, scale = 1]{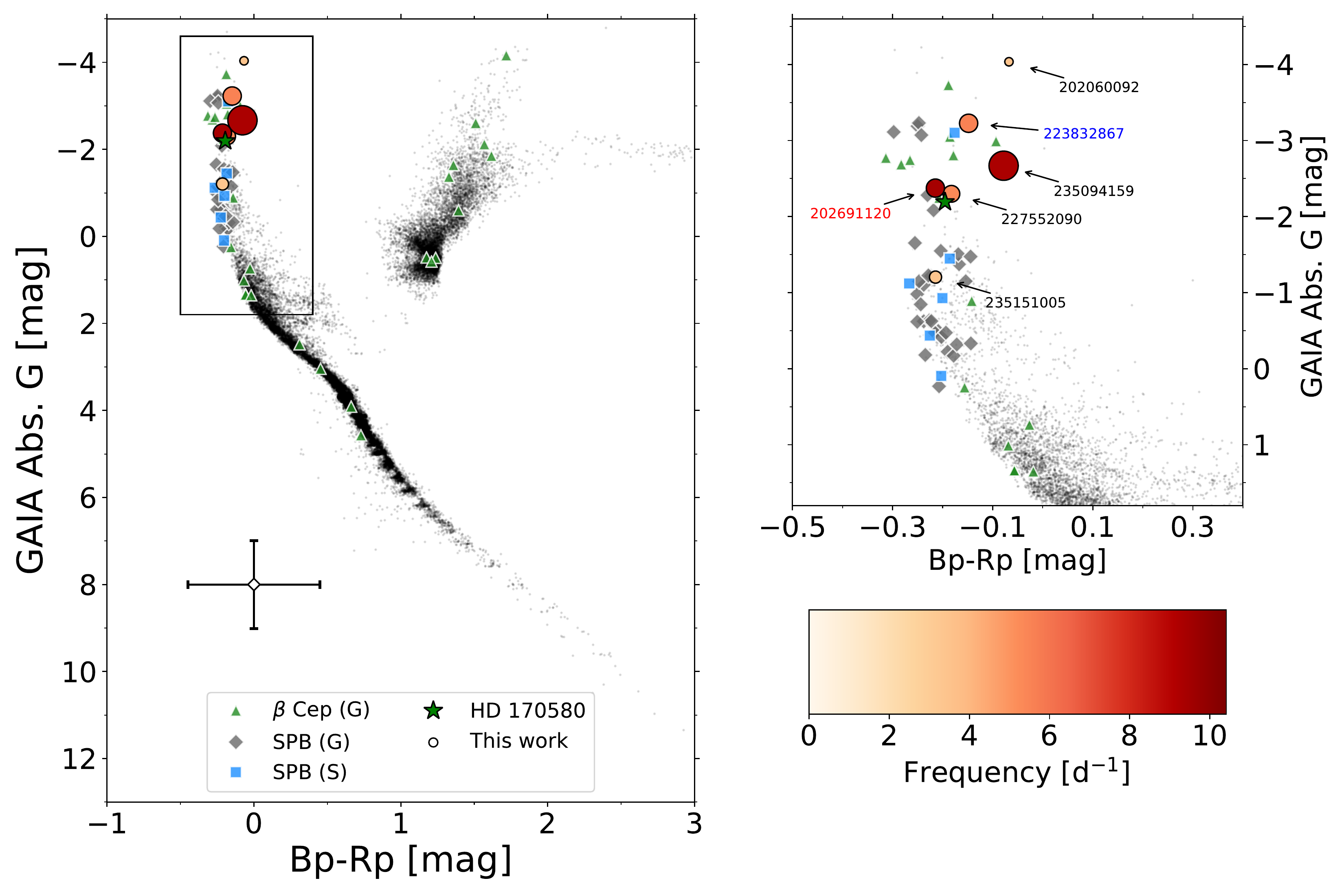}
    \caption{Observational colour-magnitude diagram for the targets presented in this work and given in Table~\ref{table:Star_List}. 
    The meaning of the symbols is shown in the legend.
    The size of the circular symbols of the stars in our sample indicates the amplitude of the dominant mode of that star, ranging between 1.9(2) and 21.3(3)~mmag, and the colour bar represents the frequency of this mode. The color of the names indicates if it is observed by \textit{TESS} in two sectors (blue), one sector (red) or not (black).
    The background points in grey denote all \textit{Kepler} objects for which Gaia-DR2 parallaxes are available. We include known $\beta$~Cep and SPB stars from ground-based photometry (G) and space-age photometry (S) as a frame of reference. The upper right frame shows a zoom of six of the eight newly discovered pulsators in the box marked on the left frame.}
    \label{fig:HRD}
\end{figure*}

We summarise the discovered variability of the eight \textit{K2} stars in Fig.~\ref{fig:HRD}, where we plot an observational colour-magnitude diagram (CMD) using the second data release, DR2 of Gaia \citep{Gaia2016, Gaia2018}. The horizontal axis denotes the difference in the apparent blue and red Gaia band pass brightness in units of magnitude corrected for reddening using values supplied by Gaia-DR2 \citep{Andrae2018}. The vertical axis denotes the absolute G-band magnitude, corrected for interstellar absorption \citep{Andrae2018}. To construct absolute magnitudes we use the distance estimations provided by \citet{Bailer-Jones2018}. The distances are obtained purely based on geometric properties (i.e. the parallax), independent of any physical properties or extinction values of individual stars. We also include an error bar to indicate the uncertainty related to the average extinction and reddening, based on the \textit{Kepler} targets monitored by Gaia. 

The size of the circles in Fig.~\ref{fig:HRD} is based on the amplitude of the dominant mode for each star and the colour bar indicates the frequency at which this maximum occurs. In this way we provide a simple representation of the diverse variability of pulsators in this small sample.
As a frame of reference we included all \textit{Kepler} stars with apparent G-band magnitude less than 13.5~mag, and having a Gaia-DR2 parallax at least four times above its error (if it is in the database in the first place). For this we made use of the gaia-kepler.fun crossmatch database created by Megan Bedell\footnote{\url{https://gaia-kepler.fun}}. 

All values from Gaia-DR2 used for our sample are summarised in Table~\ref{table:Gaia}. The location of the rest of the targets on the CMD confirms them as early-type stars, such that the spectral classification from \texttt{SIMBAD}, which we used to select our targets is at least approximately adequate for these stars. For EPIC~233986359, a parallax was not available in Gaia-DR2, and for EPIC~202929357 no reddening or extinction values were available, so those stars were not included in Fig.~\ref{fig:HRD}. 

We use the CMD to explore the relative positions and evolutionary stages of the stars in our sample with respect to known B type pulsators, due to a lack of spectroscopically derived atmospheric parameters $T_{\rm eff}$ and log~$g$. The known B type pulsators include ground-based (photometry/spectroscopy) discovered SPB and $\beta$~Cep from lists compiled by P.~De~Cat as of Jan 2004\footnote{\url{http://www.ster.kuleuven.ac.be/~peter/Bstars/}}, and space-mission photometry SPB stars as compiled in Appendix B in \citet{Szewczuk2018}. One additional star, HD~170580 \citep{Aerts2019b} is put as a reference for EPIC~235151005. In Appendix~\ref{appendix: Gaia} we verify whether the colour-magnitude diagram, in  Fig.~\ref{fig:HRD}, is a good proxy for the theoretical Hertzsprung-Russell diagram in the case where spectroscopic parameters are lacking.

For the \textit{K2} stars in our sample, additional data is needed before seismic modelling is attempted. EPIC~223832867 is visible in \textit{TESS} sectors 3 and 4 consecutively such that a time base of $\sim$54~d is available. Both EPIC~202691120 and EPIC~202929357 are observed in Sector 12, implying $\sim$27~d of additional photometry. EPIC~202060092, EPIC~233986359, EPIC~235094159, EPIC~227552090 and EPIC~235151005 are unfortunately not observed by \textit{TESS}. These additional \textit{TESS} data will be useful to add to the \textit{K2} photometry. The coherent character of the modes postulates that they have long lifetimes, as expected for $\beta$~Cep stars. We marked the stars to be observed by \textit{TESS} on Fig.~\ref{fig:HRD} in one sector for $\sim$ 27~d (red) and in two sectors for $\sim$ 54~d (blue). 

\subsection{The three new $\beta$~Cep stars}
EPIC~202929357 is the richest $\beta$~Cep in terms of the number of high amplitude resolved p-mode frequencies. Unfortunately it is not included in the CMD (Fig.~\ref{fig:HRD}), but it is similar to EPIC~202691120 with regards to the pulsational character. The location of EPIC~202691120 places it amongst previously known $\beta$~Cep and confirms the \texttt{SIMBAD} classification.  The same spatial location of EPIC~202929357 (the \textit{K2} Campaign~2 field of view) in the ecliptic implies that the reddening is approximately similar in magnitude, which would place it relatively close to EPIC~202691120. Spectroscopy is ultimately needed to confirm the classification formally.

Similarly for EPIC~235094159, the distance to this object is poorly constrained as seen in Table~\ref{table:Gaia}, $d = 6000.2_{-1548.9}^{+2673.3}$~pc, leading to an uncertain absolute G magnitude. Moreover, the harmonic series in the low frequency region points towards a pulsating B star with either ellipsoidal variability or rotational modulation. This has an effect on the observed colours as well. Spectroscopy, to identify the binary system and determine spectroscopic parameters, is needed for unambiguous classification.

\subsection{The three stars with frequency groupings}

Three stars exhibit structure in the form of frequency groupings: EPIC~202060092, EPIC~233986359 and EPIC~235151005. They cover a wide difference in colours between them (Fig.~\ref{fig:HRD}), implying that the underlying mechanism might not be related.
The underlying cause of the structures remains uncertain, as we require additional information in the form of either spectroscopy and/or additional uninterrupted photometry to increase the frequency precision. Not all frequencies can be explained from a set of parent modes.

The incompatibility of explaining the frequency group structure as combination frequencies is similar to the case of HD~170580, which was found to be an evolved star near the TAMS \citep{Aerts2019b}. Our results clearly demonstrate the necessity of long-term, continuous and high-precision photometry for studying the variability of early-type stars. Unfortunately none of the stars exhibiting frequency groupings (EPIC~202060092, EPIC~233986359, EPIC~235151005) will get observed with the \textit{TESS}.

\section{Conclusion}\label{section: conclusion}
The \textit{K2} mission and its changing field of view have allowed us to discover new variables in a sparsely populated blue part of the HRD (Fig.~\ref{fig:HRD}). The variable objects include three new $\beta$~Cep stars (EPIC~202691120, EPIC~202929357, EPIC~235094159), one known $\beta$~Cep (EPIC~223832867) for which we present additional low-amplitude frequencies, and three stars exhibiting structure in the form of distinct groups of frequencies (EPIC~202060092, EPIC~233986359 and EPIC~235151005).  

Nearly equally spaced frequency patterns were found in EPIC~227552090. Stochastic excitation induced by turbulent motions in the subsurface convection zone could be the cause \citep{Cantiello2009, Belkacem2010}, but we do not have enough information to exclude other possibilities. For the moment we cannot firmly identify any specific cause of the observed spacings. EPIC~227552090 will not be viewed by \textit{TESS} such that increasing the frequency resolution is not possible in the near future. Spectroscopic analysis of this star is a priority, as it will allow us to re-evaluate the information available from the \textit{K2} light curve. 

Future modelling of these particular stars will involve determining their the effective temperature and log~$g$ from snapshot spectroscopy.  Unfortunately not all eight stars in the sample will be observed with \textit{TESS} such that the light curves will not be extended and the frequency resolution will not improve in the near future. For those stars that will be observed, the additional data will be helpful to reconfirm the frequencies, and re-analyse \textit{TESS} photometry merged with the \textit{K2} photometry. Nonetheless, our addition of \textit{K2} high frequency pulsators to the total number of upper main sequence variable stars, and the large amplification soon by \textit{TESS}, is bound to add incrementally to the much needed calibration of stellar evolution theory of hot massive stars. With \textit{TESS}, massive star asteroseismology will certainly be getting a boost (\citealp{Pedersen2019a, Bowman2019b}).

\section*{Acknowledgements}

The research leading to these results has received funding from the European Research Council (ERC) under the European Union's Horizon 2020 research and innovation program (grant agreement No. 670519: MAMSIE).  This research has made use of the \texttt{SIMBAD} database, operated at CDS, Strasbourg, France. Some/all of the data presented in this paper were obtained from the Mikulski Archive for Space Telescopes (MAST). STScI is operated by the Association of Universities for Research in Astronomy, Inc., under NASA contract NAS5-26555. This paper includes data collected by the Kepler mission. Funding for the Kepler mission is provided by the NASA Science Mission directorate.




\bibliographystyle{mnras}




\appendix

\section{Frequency lists}\label{appendix: fl}
This appendix contains all the frequency lists extracted by pre-whitening described in Section~\ref{subsection: FA}. The first column reads the assigned ID of each pre-whitened frequency, the second column the frequency with the error on the last figure between brackets, the third column the corresponding amplitude with the error on the last figure between brackets, the fourth column the signal-to-noise of that peak. Finally the fifth column denotes any identified combinations/harmonics. In Table \ref{table:223832867_freq} the column is also used to identify peaks found previously by \citet{Pigulski2008} for EPIC~223832867.

\begin{table}
\caption{Frequency list for  EPIC~202691120 extracted by pre-whitening. The standard deviation of the residuals is: 0.063~mmag.} \label{table:202691120_freq}      
\centering          
\begin{tabular}{c c c c c}  
\hline

ID &	Frequency 	& Amplitude 	&SN & Notes \\
&[\si{\per\day}]&[mmag]&& \\
\hline

$\nu_{1}$&	9.4112(2)&	8.3(3)&	25.2& \\
$\nu_{2}$&	7.9107(2)&	6.7(2)&	28.5& \\
$\nu_{3}$&	0.3419(5)&	2.3(2)&	12.5& \\
$\nu_{4}$&	8.5187(6)&	2.1(2)&	16.1& \\
$\nu_{5}$&	8.9612(7)&	1.7(2)&	15.0& \\
$\nu_{6}$&	2.2919(7)&	1.6(2)&	9.8& \\
$\nu_{7}$&	0.0925(9)&	1.1(1)&	7.7& \\
$\nu_{8}$&	7.480(1)&	1.0(1)&	10.6& \\
$\nu_{9}$&	6.883(1)&	0.9(1)&	9.3&  \\
$\nu_{10}$&	1.307(1)&	0.8(1)&	5.8& \\
$\nu_{11}$&	1.126(1)&	0.7(1)&	5.3& \\
$\nu_{12}$&	0.986(1)&	0.7(1)&	5.3& $\nu_{6} - \nu_{10}$\\
$\nu_{13}$&	1.088(1)&	0.7(1)&	5.0& $\nu_{12}$ - $\nu_{7}$\\
\hline
\end{tabular}
\end{table}

\begin{table}
\caption{Frequency list for EPIC~202929357 extracted by pre-whitening. The standard deviation of the residuals is: 0.070~mmag. No combinations were found.} \label{table:202929357_freq}      
\centering          
\begin{tabular}{c c c c c}  
\hline

ID &	Frequency	& Amplitude 	&SN & Notes \\
& [\si{\per\day}]&[mmag]&&\\
\hline

$\nu_{1}$&	9.1640(4)&	4.8(2)&	17.4&\\
$\nu_{2}$&	10.8271(4)&	4.7(2)	&14.3&\\
$\nu_{3}$&	13.0969(6)&	3.2(2)&	15.3&\\
$\nu_{4}$&	8.0155(5)&	3.2(2)&	14.6&\\
$\nu_{5}$&	10.4413(5)&	3.1(2)&	13.8&\\
$\nu_{6}$&	8.7579(5)&	2.6(2)&	13.9&\\
$\nu_{7}$&	9.5796(6)&	2.2(1)&	12.4&\\
$\nu_{8}$&	10.1467(6)&	2.1(2)&	11.4&\\
$\nu_{9}$&	7.8876(5)&	2.1(1)&	15.3&\\
$\nu_{10}$&	12.4737(6)&	1.9(1)&	12.4&\\
$\nu_{11}$&	11.4182(8)&	1.3(1)&	8.6&\\
$\nu_{12}$&	8.8193(8)&	1.3(1)&	11.2&\\
$\nu_{13}$&	9.4718(8)&	1.3(1)&	10.6&\\
$\nu_{14}$&	11.952(1)&	1.0(1)&	7.4&\\
$\nu_{15}$&	4.0458(9)&	1.0(1)&	6.0& \\
$\nu_{16}$&	0.051(1)&	0.9(1)&	6.6& \\
$\nu_{17}$&	9.063(1)&	0.7(1)&	7.2&\\
$\nu_{18}$&	8.861(1)&	0.7(1)&	7.3&\\
$\nu_{19}$&	10.076(1)&	0.7(1)&	6.3&\\
$\nu_{20}$&	12.526(1)&	0.7(1)&	5.5&\\
$\nu_{21}$&	13.723(1)&	0.7(1)&	5.6&\\
$\nu_{22}$&	7.456(1)&	0.6(1)&	6.6&\\
$\nu_{23}$&	11.265(1)&	0.6(1)&	5.4&\\
\hline
\end{tabular}
\end{table}

\begin{table}
\caption{Frequency list for EPIC~223832867 extracted by pre-whitening.  The standard deviation of the residuals is: 0.065~mmag.
Modes identified by \citet{Pigulski2008} are marked with PP08, with their amplitude in mmag between brackets.} \label{table:223832867_freq}      
\centering          
\begin{tabular}{c c c c c}  
\hline
ID &	Frequency 	& Amplitude 	&SN & Notes \\
&[\si{\per\day}]&[mmag]&&\\
\hline

$\nu_{1}$&	5.5708(5)&	8.4(5)&	22.1& PP08 (8.6)\\
$\nu_{2}$&	5.1640(4)&	8.0(4)&	24.4& PP08 (12.0)\\
$\nu_{3}$&	5.1182(5)&	6.9(4)&	22.7& PP08 (13.9)\\
$\nu_{4}$&	4.9407(4)&	5.4(3)&	27.9& PP08 (6.6)\\
$\nu_{5}$&	0.3949(6)&	2.3(2)&	14.3& $\nu_{1}$ - $\nu_{2}$\\
$\nu_{6}$&	7.1117(7)&	1.7(2)&	12.3&\\
$\nu_{7}$&	0.3657(6)&	1.8(1)&	11.8&\\
$\nu_{8}$&	7.719(1)&	1.0(1)&	7.0&\\
$\nu_{9}$&	3.690(1)&	0.9(1)&	7.5&\\
$\nu_{10}$&	0.597(1)&	1.0(1)&	6.9&\\
$\nu_{11}$&	6.501(1)&	0.7(1)&	6.0&\\
$\nu_{12}$&	3.124(1)&	0.7(1)&	5.9&\\
$\nu_{13}$&	8.487(1)&	0.6(1)&	5.5&\\
$\nu_{14}$&	0.690(1)&	0.6(1)&	5.1&\\
$\nu_{15}$&	7.045(1)&	0.6(1)&	5.1&\\

\hline
\end{tabular}
\end{table}

\begin{table}
\caption{Frequency list for EPIC~227552090 extracted by pre-whitening. The standard deviation of the residuals is: 0.048~mmag.} \label{table:227552090_freq}      
\centering          
\begin{tabular}{c c c c c}  
\hline

ID &	Frequency 	& Amplitude 	&SN & Notes \\
&[\si{\per\day}]&[mmag]&&\\
\hline

$\nu_{1}$&	5.3131(2)&	7.0(2)&	33.2&\\
$\nu_{2}$&	6.0599(4)&	2.8(2)&	17.4&\\
$\nu_{3}$&	6.1849(6)&	1.9(1)&	13.0&\\
$\nu_{4}$&	0.4907(6)&	1.7(1)&	15.6&\\
$\nu_{5}$&	6.3551(7)&	1.5(1)&	11.0&\\
$\nu_{6}$&	7.5387(7)&	1.4(1)&	11.5&\\
$\nu_{7}$&	6.5310(7)&	1.3(1)&	10.9&$\nu_{2} + \nu_{4}$\\
$\nu_{8}$&	5.4690(7)&	1.2(1)&	11.4&\\
$\nu_{9}$&	6.4452(8)&	1.1(1)&	10.2&\\
$\nu_{10}$&	7.4972(9)&	0.9(1)&	10.2&\\
$\nu_{11}$&	0.0302(7)&	0.88(8) &	5.9& \\
$\nu_{12}$&	7.450(1)&	0.8(1)&	7.9&\\
$\nu_{13}$&	8.008(1)&	0.7(1)&	7.2&\\
$\nu_{14}$&	6.627(1)&	0.7(1)&	7.2&\\
$\nu_{15}$&	7.188(1)&	0.64(9)&	7.3&\\
$\nu_{16}$&	5.800(1)&	0.62(9)&	7.3& $\nu_{1} + \nu_{4}$\\
$\nu_{17}$&	6.149(1)&	0.60(9)&	6.9&\\
$\nu_{18}$&	5.644(1)&	0.54(9)&	6.8&\\
$\nu_{19}$&	10.782(1)&	0.47(8)&	7.0&\\
$\nu_{20}$&	8.831(1)&	0.45(8)&	5.7&\\
$\nu_{21}$&	8.185(2)&	0.43(8)&	5.4&\\
$\nu_{22}$&	10.628(2)&	0.43(8)&	6.9& 2$\nu_{1}$\\
$\nu_{23}$&	0.108(1)&	0.42(8)&	5.2&\\
$\nu_{24}$&	0.338(2)&	0.39(8)&	5.2&\\

\hline
\end{tabular}
\end{table}

\begin{table}
\caption{Frequency list for EPIC~233986359 extracted by pre-whitening. The standard deviation of the residuals is: 0.1606~mmag.} \label{table:233986359_freq}      
\centering          
\begin{tabular}{c c c c c}  
\hline

ID &	Frequency 	& Amplitude 	&SN & Notes \\
&[\si{\per\day}]&[mmag]&&\\
\hline

$\nu_{1}$ (fg1)&	2.7057(5)&	8.2(5)&	19.7& \\
$\nu_{2}$ (fg1)&	2.7495(4)&	7.1(4)&	18.6&\\
$\nu_{3}$ (fg2)&	5.6519(9)&	3.4(4)&	14.0&\\
$\nu_{4}$ (fg2)&	5.7137(9)&	3.1(4)&	12.8&  \\
$\nu_{5}$ (fg0)&	0.066(1)&	2.4(3)&	9.1& $\nu_{3} - \nu_{4}$\\
$\nu_{6}$ (fg2)&	5.267(1)&	2.3(3)&	9.0&\\
$\nu_{7}$ (fg2)&	5.381(1)&	2.3(3)&	7.9&\\
$\nu_{8}$ (fg2)&	5.417(1)&	2.0(3)&	8.0& 2$\nu_{1}$\\
$\nu_{9}$ (fg0)&	0.043(1)&	1.9(3)&	5.3& $\nu_{2} - \nu_{1}$\\
$\nu_{10}$ (fg0)&	0.132(1)&	1.7(3)&	5.6& 2$\nu_{5}$\\
$\nu_{11}$ (fg2)&	5.327(2)&	1.6(3)&	7.1&\\
$\nu_{12}$ (fg2)&	5.134(2)&	1.6(3)&	6.7&\\
$\nu_{13}$ (fg1)&	3.088(1)&	1.5(3)&	5.2&\\

\hline
\end{tabular}
\end{table}

\begin{table}
\caption{Frequency list for EPIC~235094159 extracted by pre-whitening.
The standard deviation of the residuals is 0.085~mmag.} \label{table:235094159_freq}      
\centering          
\begin{tabular}{c c c c c}  
\hline

ID &	Frequency 	& Amplitude 	&SN & Notes \\
&[\si{\per\day}]&[mmag]&&\\
\hline

$\nu_{1}$&	9.3090(1)&	21.3(3)&	47.1& \\
$\nu_{2}$&	8.7557(3)&	7.7(3)&	28.9&\\
$\nu_{3}$&	0.0381(4)&	5.0(3)&	15.9&\\
$\nu_{4}$&	11.9416(6)&	3.7(3)&	16.4&\\
$\nu_{5}$&	9.1468(5)&	3.2(2)&	14.8&\\
$\nu_{6}$&	0.0838(6)&	2.6(2)&	9.6&  2$\nu_{3}$\\
$\nu_{7}$&	6.5384(6)&	2.5(2)&	15.3&\\
$\nu_{8}$&	11.4840(7)&	2.2(2)&	10.8&\\
$\nu_{9}$&	9.5727(8)&	1.9(2)&	10.1&\\
$\nu_{10}$&	0.1073(8)&	1.7(2)&	7.7& 3$\nu_{3}$\\
$\nu_{11}$&	7.3560(9)&	1.5(2)&	9.9&\\
$\nu_{12}$&	8.7034(9)&	1.5(2)&	9.1&\\
$\nu_{13}$&	13.672(1)&	1.3(2)&	9.0&\\
$\nu_{14}$&	10.299(1)&	1.3(2)&	7.5& \\
$\nu_{15}$&	10.927(1)&	1.3(2)&	7.6&\\
$\nu_{16}$&	1.918(1)&	1.2(2)&	7.1&\\
$\nu_{17}$&	13.290(1)&	1.1(2)&	7.8&\\
$\nu_{18}$&	1.562(1)&	1.0(2)&	6.2&\\
$\nu_{19}$&	16.568(1)&	1.0(2)&	9.6&\\
$\nu_{20}$&	9.362(1)&	1.0(1)&	6.4&\\
$\nu_{21}$&	8.099(1)&	0.9(2)&	7.2&\\
$\nu_{22}$&	8.232(1)&	0.9(1)&	6.9&\\
$\nu_{23}$&	4.362(1)&	0.9(1)&	7.9&\\
$\nu_{24}$&	12.776(1)&	0.9(1)&	6.8&\\
$\nu_{25}$&	14.736(1)&	0.9(1)&	7.9&\\
$\nu_{26}$&	9.125(1)&	0.9(1)&	6.3&\\
$\nu_{27}$&	11.898(1)&	0.9(1)&	6.3&\\
$\nu_{28}$&	0.215(1)&	0.8(1)&	5.5& \\
$\nu_{29}$&	1.293(1)&	0.8(1)&	5.5&\\
$\nu_{30}$&	1.045(1)&	0.8(1)&	5.6&\\
$\nu_{31}$&	12.411(1)&	0.8(1)&	6.4&\\
$\nu_{32}$&	0.350(1)&	0.8(1)&	5.4& 9$\nu_{3}$\\
$\nu_{33}$&	0.304(1)&	0.8(1)&	5.3& 8$\nu_{3}$\\
$\nu_{34}$&	0.938(1)&	0.8(1)&	5.2&\\
$\nu_{35}$&	10.582(1)&	0.7(1)&	5.7&\\
\hline
\end{tabular}
\end{table}

\begin{table}
\caption{Frequency list for EPIC~235151005 extracted by pre-whitening. The standard deviation of the residuals is 0.032~mmag.} \label{table:235151005_freq}      
\centering          
\begin{tabular}{c c c c c}  
\hline

ID &	Frequency 	& Amplitude 	&SN & Notes \\
&[\si{\per\day}]&[mmag]&&\\
\hline

$\nu_{1}$ (fg2)&	3.3812(3)&	3.7(2)&	31.6& $\nu_{4}+\nu_{15}$, $\nu_{3}+\nu_{15}$\\
$\nu_{2}$ (fg2)&	3.1203(8)&	1.6(2)&	15.1& 2$\nu_{9}$, $\nu_{11}+\nu_{16}$\\
$\nu_{3}$ (fg1)&	1.9134(7)&	1.54(1)&14.3& \\
$\nu_{4}$ (fg1)&	1.884(1)&	1.0(1)&	11.3&\\
$\nu_{5}$ (fg2)&	3.152(1)&	0.9(1)&	10.1&\\
$\nu_{6}$ (fg2)&	3.030(1)&	0.8(1)&	9.2& $\nu_{9}+\nu_{16}$\\
$\nu_{7}$ (fg0)&	0.031(1)&	0.8(1)&	9.1& $\nu_{3} - \nu_{4}$, $\nu_{15}-\nu_{16}$  \\
$\nu_{8}$ (fg2)&	3.182(1)&	0.7(1)&	9.3&\\
$\nu_{9}$ (fg1)&	1.565(1)&	0.7(1)&	9.1& 1/2$ \nu_{2}$\\
$\nu_{10}$ (fg0)&	0.302(1)&	0.6(1)&	7.5&\\
$\nu_{11}$ (fg1)&	1.635(1)&	0.56(9)&	7.5&\\
$\nu_{12}$ (fg0)&	0.352(1)&	0.54(9)&	7.7& $\nu_{3} - \nu_{9}$ \\
$\nu_{13}$ (fg2)&	3.333(2)&	0.42(9)&	7.2&\\
$\nu_{14}$ (fg2)&	3.797(2)&	0.40(8)&	6.3& $\nu_{3} + \nu_{4}$\\
$\nu_{15}$ (fg1)&	1.496(2)&	0.39(8)&	5.6& \\
$\nu_{16}$ (fg1)&	1.470(2)&	0.38(8)&	5.9&\\
$\nu_{17}$ (fg2)&	3.411(2)&	0.35(7)&	5.8& $\nu_{3}+\nu_{15}$\\
$\nu_{18}$ (fg2)&	3.826(2)&	0.34(8)&	6.0& 2$\nu_{3}$\\
$\nu_{19}$ (fg0)&	0.219(2)&	0.31(7)&	5.0& \\
$\nu_{20}$ (fg2)&	3.492(2)&	0.30(7)&	5.3& $\nu_{3}+\nu_{9}$\\
$\nu_{21}$ (fg2)&	3.056(2)&	0.29(7)&	5.0& $\nu_{9}+\nu_{15}$\\
\hline
\end{tabular}
\end{table}

\section{Gaia}\label{appendix: Gaia}
\begin{table*}
\caption{Gaia-DR2 values for the stars in our sample retrieved from \url{http://gea.esac.esa.int/archive/}. The last three columns are distances from \citet{Bailer-Jones2018}. EPIC~233986359 is not in Gaia-DR2 and therefore not included in the table.}\label{table:Gaia}     
\begin{threeparttable}
\centering      
\begin{tabular}{c c c c c c c c c c}     
\hline\hline       
EPIC~ID & Gaia ID & G mean mag. & G absorption & BP-RP & E(BP-RP) & Parallax & Distance & Lower bound & Upper bound\\ 
&&[mag]&[mag]& [mag]& [mag]& [arcsec] & [pc] & [pc] & [pc]\\
\hline
202060092 & 3377246313718987392	&	8.956&	1.676&	0.806&	0.872	&	0.517&	1832.7&	1675.4&	2021.7\\
202691120& 6032148702447357952	&	10.232&	1.256&	0.425&	0.640	&	0.512&	1862.6&	1708.9&	2046.0\\
202929357& 6042284859603154944&		10.873&	NA$^{*}$& 	0.236&		NA$^{*}$&		0.333&	2790.2&	2410.0&	3306.1\\
223832867& 4065715861804605696	&	8.900&	1.407&	0.532&		0.680&		0.698&	1392.5&	1265.5&	1547.3\\
227552090& 4096294139196441216	&	9.752&	1.338&	0.491&		0.674&		0.701&	1389.0&	1246.0&	1568.1\\
235094159& 4112611952506129408&		12.369&	1.147&	0.501&		0.579&		0.126&	6000.2&	4451.2&	8673.5\\
235151005& 4111320477373506304&	8.642&	1.651&	0.621&	0.835&		2.277&	435.3&	416.4&	456.1\\
\hline

\end{tabular}
\begin{tablenotes}
\small
\item $^{*}$ indicates that the extinction in the G-band is poorly estimated and the value untrustworthy \citep{Andrae2018}.
\end{tablenotes}
\end{threeparttable}
\end{table*}

In this appendix we summarise the Gaia data used for the colour-magnitude diagram in Fig.~\ref{fig:HRD} in Table~\ref{table:Gaia}. The last three columns are values from \citet{Bailer-Jones2018}.
To confirm the validity of the diagram in Fig.~\ref{fig:HRD} we use approximate estimates of the effective temperature from spectral type and the Gaia-DR2 parameters to derive the luminosities. We follow the method of \citet{Pedersen2019b} to achieve this and provide only a summary.

The method starts from the observed apparent magnitudes in different filters and applies bolometric corrections,
\begin{equation}
m_{\rm bol} = m_{S_{\lambda}} + BC_{S_{\lambda}} - A_{S_{\lambda}}, \end{equation}
with $m_{\rm bol}$ the apparent bolometric magnitude, $m_{S_{\lambda}}$ the apparent magnitude measured in photometric passband with transmission curve $S_{\lambda}$, and $BC_{S_{ \lambda}}$ and $A_{S_{\lambda}}$ the bolometric correction and extinction for this transmission curve $S_{\lambda}$. Bolometric corrections $BC_{S_{\lambda}}$ have been computed for Gaia passbands but are restricted to the temperature range of $T_{\rm eff}\in$~[2600-8000K] (see e.g. \citealt{Andrae2018}). Therefore \citet{Pedersen2019b} computed $BC_{S_{\lambda}}$ for the temperature regime relevant for SPB and $\beta$~Cep stars.

The extinction values $A_{S_{\lambda}}$ are procured from the reddening values $E(B-V)$ of each individual star and the reddening law,
\begin{equation}
A_{S_{\lambda}} = R_{S_{\lambda}} E(B-V),
\end{equation}
with $R_{S_{\lambda}}$ the ratio of total to selective reddening in the passband with transmission $S_{\lambda}$. \citet{Pedersen2019b} obtain $E(B-V)$ from 3D dust maps by \citet{Green2018}, and derive $R_{S_{\lambda}}$ vs. $\lambda$ using the York's extinction solver \citep{McCall2004} assuming the \citet{Fitzpatrick1999} reddening law. This finally yields the apparent bolometric magnitude $m_{\rm bol}$ for each star.

Using the distances from Gaia-DR2 \citep{Gaia2016, Gaia2018}, see Table~\ref{table:Gaia}, as calculated by \citet{Bailer-Jones2018}, the luminosities are then retrieved through,
\begin{equation}
\log~L_{\star}/L_{\odot} = -0.4 (m_{\rm bol} - 5 \log~d + 5 - M_{\rm bol,\odot}),
\end{equation}
where $d$ is the distance measured in parsec and $M_{\rm bol,\odot}$ the absolute bolometric magnitude of the Sun.

The lack of availability of spectroscopy forces us to derive the effective temperatures from spectral type-effective temperature calibrations, e.g. as provided by \citet{Dejager1987} for B-type stars. The authors of this work collected the most recent and complete (in their time) sets of tabular values of effective temperature and spectral type. The sets were then used to derive relations where log~$T_{\rm eff}$ and  log~$L_{\star}$ are written as Chebyhev polynomials depending on two continuous analytical variables $s$ (connected to spectral class) and $b$ (connected to luminosity class), such that log~$T_{\rm eff}(s,b)$ and log~$L_{\star}(s,b)$. The authors provided tables from which, given a spectral type and luminosity class, the effective temperature is procured. 
The 1$\sigma$, which is 0.012 per unit weight for log~$T_{\rm eff}$, is given as the approximate error on any  log~$T_{\rm eff}$ derived from the spectral type \citep{Dejager1987}. This is the uncertainty that we will also use.

For EPIC~202060092, estimates of the effective temperature and surface gravity are provided by \citet{Buysschaert2015} so we use that value instead. For EPIC~20269110 and EPIC~202929357 the spectral type classification was assigned OB$^{-}$ by \citet{Drilling1995}. We noted in Section~\ref{sec:202691120} that in their scheme, OB was assigned to anything that appears earlier than B6 but that late B-type or early A-type supergiants may imitate OB behaviour \citep{Nassau1960}. This covers a large range in temperatures and luminosities. To cover the extreme case that the stars may be late B-type or early A we assume $T_{\rm eff}=15000\pm 5000$K which covers the mid to late B range. For all stars, expect EPIC~202060092, we assume a surface gravity log~$g=4.0\pm 0.5$, which approximately reflects the main sequence in this mass range.  We summarise the relevant parameters in Table~\ref{table:StarParam}.
The errors on the luminosity are derived by means of error propagation from the uncertainty on the distance, and temperature, as well as bolometric correction. See \citet{Pedersen2019b} for details.

\begin{figure}
	\includegraphics[width=1\columnwidth, scale = 1]{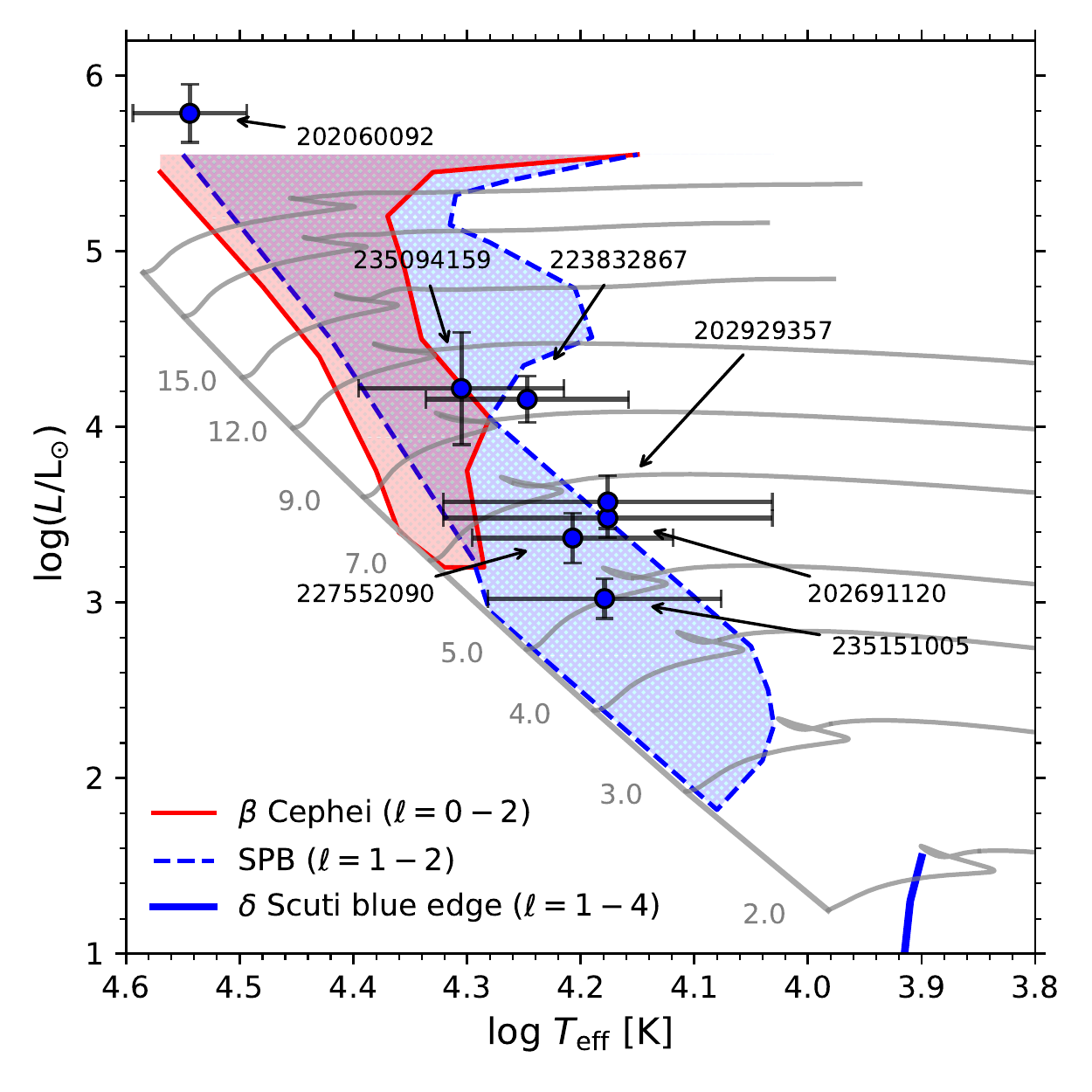}
    \caption{Theoretical Hertzsprung-Russell diagram using effective temperatures from \citet{Dejager1987} and luminosities derived using Gaia-DR2 and the statistical model by \citet{Pedersen2019b}.The grey evolutionary tracks are by \citet{Johnston2019}, the massive star instability strips from \citet{Walczak2015}, and the $\delta$~Scuti blue edge is from \citet{Dupret2004}.}
    \label{fig:tHRD}
\end{figure}

The final results of this exercise are summarised in Fig.~\ref{fig:tHRD}. On this figure we show the positions of the stars in terms of luminosity and effective temperature. For illustrative purposes, we include evolutionary tracks computed by \citet{Johnston2019} with the MESA stellar evolution code \citep{Paxton2011, Paxton2013, Paxton2015, Paxton2018}. These are non-rotating tracks with Z = 0.014, an initial hydrogen abundance X$_{\rm ini} = 0.71$, diffuse exponential overshooting parameter $f_{\rm ov}=0.020$, the \citet{Asplund2009} heavy element mixture, and the standard MESA opacity tables that incorporate different sources \citep{Paxton2011}.
Additionally we include two approximate instability strips: the $\beta$~Cephei and SPB instability regions, as computed by \citet{Walczak2015}. These were calculated using a combination of the MESA stellar evolution code and the Dziembowski linear, non-adiabatic pulsation code \citep{Dziembowski1977}, for Z = 0.015, X$_{\rm ini} = 0.70$, exponential overshooting parameter $f_{\rm ov}=0.020$, the \citet{Asplund2009} solar mixture, and OPAL opacity tables. We also include the $\delta$~Scuti blue edge for $\ell = 1-4$ from \citet{Dupret2004}, for solar metallicity. We emphasise that the strips and the blue $\delta$~Scuti edge are included to guide the eye, and are not meant as a comparison between theory and observations. The $\beta$~Cep strip is sensitive to many parameters, i.e. metallicity, given that the modes are excited by a heat mechanism \citep{Szewczuk2017}.

The relative positions of the stars are very similar to the positions in Fig.~\ref{fig:HRD}, demonstrating that the colour-magnitude diagram is a suitable proxy for the theoretical Hertzsprung-Russell diagram whenever spectroscopic parameters are not available. The error bars on the temperatures are high, as expected, and this is also propagated into the luminosity. Fig.~\ref{fig:tHRD} indicates that the stars are most likely unstable to low and/or high frequency oscillations.

\begin{table*}
\caption{Assumed stellar parameters, reddening values and derived luminosities following using the model by \citet{Pedersen2019b}.}\label{table:StarParam}     
\begin{threeparttable}
\centering      
\begin{tabular}{c c c c |c c }     
\hline\hline       
EPIC~ID & SpT& log~$T_{\rm eff}$ & log~$g$  & $E(B-V)$ &log~$L$ \\ 
&&[K]&[cm/s$^{2}$]&\citet{Green2018}& [$L_{\odot}$]\\
\hline
202060092$^{1}$ & O9V:p	& $4.5 \pm 0.1$&	$4.5\pm0.5$& $1.22\pm0.02$ &  $6.0\pm0.2$\\
202691120$^{2}$& OB$^{-}$	&	$4.2\pm0.1$&	$4.0\pm0.5$& $0.47\pm0.01$ &  $3.5\pm0.1$\\
202929357$^{2}$& OB$^{-}$ &	 $4.2\pm0.1$&	$4.0\pm0.5$& $0.44\pm0.00001$ &  $3.6\pm0.2$ \\
223832867& B2Ib/II	&	$4.3\pm0.1$&	$4.0\pm0.5$& $0.60\pm0.01$ &  $4.2\pm0.1$\\
227552090& B3Ib	& $4.2\pm0.1$ &	$4.0\pm0.5$& $0.33\pm0.01$ &  $3.3\pm0.1$\\
235094159& B2III &	$4.3\pm0.1$& $4.0\pm0.5$& $0.62\pm0.13$ &  $4.2 \pm 0.3$\\
235151005& B5III & $4.2\pm 0.1$& $4.0\pm0.5$ & $0.56\pm 0.0008$ &  $3.0\pm0.1$\\
\hline

\end{tabular}
\begin{tablenotes}
\small
\item (1) Values from \citet{Buysschaert2015}.
\item (2) The spectral type classification is vague, therefore the temperature cannot be retrieved from \citet{Dejager1987}.
\end{tablenotes}
\end{threeparttable}
\end{table*}

\bsp	
\label{lastpage}
\end{document}